\documentclass[twocolumn,showpacs,amsmath,amssymb]{revtex4}

\usepackage{graphicx}
\usepackage{dcolumn}
\usepackage{bm}

\begin{document}
\title{Particle acceleration by circularly and elliptically polarised dispersive Alfven waves
in a transversely inhomogeneous plasma in the inertial and kinetic regimes}
\author{D. Tsiklauri}
\affiliation{Astronomy Unit, School of Physics and Astronomy, 
Queen Mary University of London, Mile End Road, London, E1 4NS, United Kingdom}
\date{\today}
\begin{abstract}
{Dispersive Alfven waves (DAWs)  offer, an alternative to magnetic reconnection,
opportunity to accelerate solar flare particles in order to alleviate
the problem of delivering flare energy to denser parts of the solar atmosphere
to match X-ray observations.}  
{Here we focus on the effect of DAW polarisation, left, right, circular and elliptical,
in the different regimes inertial and kinetic, aiming to study these effects on the 
efficiency of particle acceleration.}
{We use 2.5D particle-in-cell simulations to study how the particles are accelerated when 
DAW, triggered by a solar flare, propagates
in the transversely inhomogeneous plasma that mimics solar coronal loop.}
{(i) In the inertial regime, fraction of accelerated electrons (along the magnetic field),
in the density gradient regions is $\approx 20\%$ by the time when DAW develops
three wavelengths and  is increasing to 
$\approx 30\%$ by the time when DAW develops
thirteen wavelengths.
In all considered cases ions are heated in the transverse to
the magnetic field direction and fraction of heated particles is $\approx 35 \%$.
(ii) The case of
right circular, left and right elliptical polarisation DAWs, with the
electric field in  the non-ignorable transverse direction exceeding several times that of
in the ignorable direction, produce more pronounced parallel electron beams (with larger maximal
electron velocities) and transverse ion beams
in the ignorable direction. 
In the inertial regime such polarisations yield the
fraction of accelerated electrons  $20\%$.
In the kinetic regime this increases to $35\%$.
(iii) The parallel electric field that is generated in the density inhomogeneity
regions is independent of the electron-ion mass ratio and stays of the order $0.03\omega_{pe}c m_e /e$
which for solar flaring plasma parameters exceeds Dreicer electric field by eight orders of magnitude.
(iv) Electron beam velocity has the phase velocity of the DAW.
Thus electron acceleration is via Landau damping of DAWs.
For the Alfven speeds of $V_A=0.3c$ the considered mechanism can accelerate electrons to energies circa 20 keV.
(v) The increase of mass ratio from $m_i/m_e=16$ to
73.44 increases fraction of accelerated electrons from $20\%$
to $30-35\%$ (depending on DAW polarisation).
For the mass ratio $m_i/m_e=1836$ the fraction of accelerated electrons
would be $>35\%$. 
(vi) DAWs generate significant
density and temperature perturbations that are located in the density gradient regions. }
{DAWs propagating in the transversely inhomogeneous plasma can effectively accelerate electrons
along the magnetic field and heat ions across it.}
\end{abstract}	

\pacs{52.35.Hr; 52.35.Qz; 52.59.Bi; 52.59.Fn; 52.59.Dk; 41.75.Fr; 52.65.Rr}


\maketitle

\section{Introduction}

Charged particles, accelerated to the speeds exceeding local thermal velocity, play a 
vital role in numerous laboratory, space, solar and astrophysical interdisciplinary processes. 
For example, in space plasmas, the 
Van Allen radiation belts form a torus of energetic charged plasma particles around Earth, 
which is held in place by Earth's magnetic field. These accelerated, charged particles come 
from cosmic rays. The radiation belts pose serious hazard to the space technology and future 
manned space travel. The outer electron radiation belt is mostly produced by the inward 
radial diffusion \cite{2004GeoRL..3108805S} and local 
acceleration \cite{2005Natur.437..227H} due to transfer of energy from whistler 
mode plasma waves to radiation belt electrons. Dispersive Alfven waves (DAWs) are known 
to accelerate electrons in the Earth auroral zone \cite{2000SSRv...92..423S}. In solar coronal plasmas, solar 
flare models involve magnetic reconnection which converts the magnetic energy into heat 
and kinetic energy of accelerated particles. About 50-80\% of the energy released during 
solar flares is converted into the energy of accelerated particles \cite{2004JGRA..10910104E}. Thus, knowledge 
of the details of particle acceleration via magnetic reconnection (which is deemed a major 
mechanism operating solar and stellar flares) is of great importance. This is of direct 
relevance to the solar activity that affects humankind via hazardous phenomena such as 
solar flares and energetic particles. In astrophysical plasmas, accelerated particles 
play a major role in explaining the existence of soft excess emission originating from 
clusters of galaxies that is of fundamental cosmological importance \cite{2008SSRv..134..207P}. Also, 
multi-wavelength observations of Active Galactic Nuclei have now provided evidence 
that they are site of production of high temperature plasmas ($kT > 100$ MeV) with strong 
tails of accelerated particles. The generation of hot, relativistic plasmas and 
accelerated particles is a challenging problem both in the context of Active 
Galactic Nuclei and the origin of high-energy cosmic rays \cite{1984AdSpR...4..345F}.
In laboratory plasmas, in the case of ions, plasma regimes in which enhanced fast-ion 
transport occurs because of strongly driven energetic particle models (EPMs) should be avoided in 
order to reach reactor relevant performance and 
prevent first wall damage \cite{2006PPCF...48B..15Z}. The investigation 
of the transition from regimes characterised by weakly driven modes (such as Alfven Eigenmodes, 
e.g. toroidal Alfven eigenmodes, TAEs) to regimes dominated by strongly driven modes (like EPMs) 
and the determination of the related threshold are then of crucial importance for burning plasma 
operations \cite{2009NucFu..49g5024V}. In the case 
of electrons, e.g. in the laser driven inertial confinement fusion, 
accelerated electrons carry away a lot of energy pumped by the laser. Also, due to the laser's 
strong coupling with hot electrons, premature heating of the dense plasma (ions) is problematic 
and fusion yields are low. Significant improvements in the modelling of these phenomena have 
been achieved \cite{2010PhPl...17e6702K}. 
Edge Localised Modes (ELMs) are MHD instabilities destabilised by the pressure 
gradient in the H-mode Tokamak edge pedestal. ELMs produce losses of up to 10\% plasma energy 
in several 100 micro-seconds. ELMs are major concern for the operation of ITER. Thus, ELM control 
is essential. ELMs are the focus of increasing attention by the edge physics community because 
of the potential impact that the large divertor heat pulses due to ELMs would have on the 
divertor design of future high power tokamaks such as ITER. Ref.\cite{1997JNuM..241..182H} reviews what is known 
about ELMs, with an emphasis on their effect on the scrape-off layer and divertor plasmas. 
ELM effects have been measured in the ASDEX-U, C-Mod, COMPASS-D, DIlI-D, JET, JFT-2M, JT-60U 
and TCV tokamaks, and were reported by Ref.\cite{1997JNuM..241..182H}. ELMs are known to produce super-thermal 
particle fluxes that are damaging for Tokamak wall materials.

At the beginning of the previous paragraph, interdisciplinary nature of the role 
of accelerated particles was emphasised. In all of the above mentioned fields of 
research, the mechanisms that lead to accelerated particles are in fact common, 
and can be split into three broad classes: (1) Electric field acceleration; 
(2) Shock acceleration and (3) Stochastic acceleration. 

In the electric field acceleration mechanism,  Ref.\cite{1959PhRv..115..238D} considered dynamics 
of electrons under the action of two effects: the parallel (to background magnetic field) 
electric field and friction between electrons and ions. The equation describing 
electron motion along the magnetic field is  
$m_e \dot{v}_d = eE - \nu_{ei} m_e v_d$, where $v_d$ is the electron 
drift velocity, $\nu_{ei}$ is the electron-ion collision frequency and dot denotes time derivative. 
In the limiting case of $v_d  << v_{thermal}$ and the steady state regime,
the equation of motion  allows
one to obtain the expression for Spitzer resistivity. In the limit of $v_d  > v_{thermal}$, the steady state 
solution is not possible. This is because when the right hand side of the equation is positive 
i.e. when $eE >\nu_{ei} m_e v_d$, one has electron acceleration. The latter leads to 
the so-called run-away regime. 
The acceleration due to the parallel electric field leads to an increase 
in the electron drift speed $v_d$, in turn, this leads to a 
decrease in $\nu_{ei}$ because $\nu_{ei} \propto v_d^{-3}$. 
Therefore, when the electric field 
exceeds the critical value $E_D=n_e e^3 \ln \Lambda/(8 \pi \varepsilon_0^2k_BT)$ 
(so-called Dreicer electric field), the faster electron motion leads to the 
decrease in electron-ion friction, which in turn results in even faster motion. 
As a result the run-away regime is reached. For electric fields less than the Dreicer field, 
the rate of acceleration is less than the rate of collision losses and only a small 
fraction of the particles can be accelerated into a non-thermal tail of energy $E < LeE_D$ \cite{2008SSRv..134..207P}, 
where $L$ is the acceleration lengthscale. Super-Dreicer fields, which seem to be present 
in many simulations of reconnection \cite{2006ApJ...644L.145C,2005ApJ...618L.111Z}, accelerate particles at a rate that is 
faster than the collision or thermalisation time. This can lead to a runaway and an 
unstable electron distribution which, as shown theoretically, by laboratory experiments 
and by the simulations \cite{2006ApJ...644L.145C,2005ApJ...618L.111Z}, 
will give rise to plasma waves or turbulence \cite{1970PhRvL..25..706B,1985ApJ...293..584H}.

In shock acceleration mechanism, which is also called first-order Fermi acceleration, 
the fractional momentum gain, $\Delta p/p$, per shock crossing is linear in the velocity ratio, $u_{sh}/v$ , 
where $u_{sh}$ and $v$ are shock and particle speeds respectively. 
In stochastic acceleration mechanism, which is also called second-order Fermi acceleration, 
particles of velocity $v$ moving along magnetic field lines with a pitch 
angle $\cos\mu$ undergo random scattering by moving agents with a velocity $u$. Because the 
head collisions (that are energy gaining) are more probable than trailing collisions 
(that are energy losing), on average, the particles gain momentum at a rate proportional 
to $(u / v)^2 D_{\mu\mu}$, where $D_{\mu\mu}$ is the pitch angle diffusion rate. Note that here dependence 
on the velocity ratio is quadratic (hence the name second order) \cite{2008SSRv..134..207P}.

In this work we focus on the electric field acceleration mechanism. 
In particular when the electric field is provided by the DAWs. 
In magnetohydrodynamic (MHD) regime Alfven waves (AWs) cannot provide parallel electric fields, therefore 
cannot accelerate particles via the electric field acceleration mechanism. After it 
was realised \cite{1970PhFl...13..440S} that in the kinetic regime (in which plasma is treated as a 
collection of charged particles, rather than a fluid as in MHD) AWs with parallel 
electric fields are possible, a significant effort has been undertaken to study these waves. 
In laboratory plasmas, an externally applied oscillating magnetic field (at a frequency near 
1 MHz for typical Tokamak parameters) was found to resonantly mode convert to a kinetic AW, 
whose perpendicular wavelength is comparable to the ion gyroradius \cite{1976PhFl...19.1924H}. 
The kinetic AW, while it propagates into the higher-density side of the plasma 
after mode conversion, dissipates due to both linear and nonlinear processes 
and heats the plasma. If a magnetic field of 50 G effective amplitude is applied, 
approximately 10 MJ m$^{-3}$ of energy can be deposited in 1 sec into the plasma. 
In space plasmas, surface waves, excited either by a MHD plasma instability or 
an externally applied impulse, were shown to convert in a resonant mode to a 
kinetic AW having a wavelength comparable to the ion gyroradius in the direction 
perpendicular to the magnetic field. The kinetic AW was found to have a component 
of its electric field in the direction of the ambient magnetic field and can 
accelerate plasma particles along the field line. A possible relation between 
this type of acceleration and the formation of auroral arcs was discussed \cite{1976JGR....81.5083H}. 
In solar coronal plasmas, it was found that an Alfven surface wave can be excited 
by 300 s period chromospheric oscillations which then resonantly excite kinetic AWs. 
Because of their kinetic structure, the latter waves were found to dissipate their energy efficiently 
\cite{1978ApJ...226..650I}.

The parallel electric field is also produced when low frequency ($\omega < \omega_{ci}$, where $\omega_{ci}=eB/m_i$ 
is the ion cyclotron frequency) 
dispersive AW has a wavelength perpendicular to the background magnetic field comparable to any of 
the kinetic spatial scales such as: ion gyroradius at electron temperature, 
$\rho_s=\sqrt{k_B T_e/m_i}/ \omega_{ci}$, ion thermal 
gyroradius, $\rho_i=\sqrt{k_B T_i/m_i}/ \omega_{ci}$, \cite{1976JGR....81.5083H} or to electron inertial length 
$\lambda_e = c/ \omega_{pe}$ \cite{1979JGR....84.7239G}. The central issue in understanding 
what physical effect(s) can provide the parallel electric field can be understood based 
on the two-fluid theory of dispersive AWs \cite{2000SSRv...92..423S}. 
The electron equation of motion, $\partial \vec{v}_e /\partial t +(\vec{v}_e \cdot \nabla)\vec{v}_e=
-(e/m_e)(\vec{E}+\vec{v}_e\times \vec{B})-\nabla \cdot {P}_e/(m_e n_e)$,  indicates 
that in the linear regime, $E_\parallel$ can be supported either by the electron inertia term, $\partial \vec{v}_e /\partial t$,
or the 
electron pressure tensor term, $\nabla \cdot {P}_e/(m_e n_e)$. 
Dispersive Alfven waves are subdivided into Inertial AWs or Kinetic AWs 
depending on the relation between the plasma $\beta$ and  electron/ion mass ratio $m_e/m_i$.
 When $\beta \ll m_e/m_i$ (i.e. when Alfven speed is much greater than electron and ion thermal speeds, 
 $V_A \gg v_{th,i}, v_{th,e}$) dominant mechanism for sustaining $E_\parallel$ is the parallel electron inertia 
 (that is why such waves are called inertial AWs). When $\beta \gg m_e/m_i$, (i.e. when  $V_A \ll v_{th,i}, v_{th,e}$) 
 then clearly thermal effects become important. Thus, the dominant mechanism for 
 sustaining $E_\parallel$ is the parallel electron pressure gradient (that is why such waves 
 are called kinetic AWs - kinetic motion of electrons is the source of pressure). 
 In the MHD regime, it has been known \cite{1983A&A...117..220H} that the phase-mixing mechanism, in which 
 an AW moves along the density inhomogeneity that is transverse to the background magnetic 
 field, can provide a fast dissipation mechanism. Fast in a sense that AW amplitude decrease, 
 as a function of travel distance, $x$, along wave propagation direction, is a much faster 
 function of $x$, $B_\perp(x)=\exp[(-\eta V_A ^{\prime 2} k^2/6 V_A^3)x^3]$, 
 compared to the usual resistive damping law $B_\perp(x)=
 \exp[(-\eta k^2/V_A)x]$, where $\eta$ is the plasma resistivity and 
 $V_A ^{\prime}$ denotes spatial derivative of the Alfven speed
 in the transverse to the background magnetic field direction \cite{1983A&A...117..220H}.
The fast dissipation occurs because the phase mixing effect
leads to the build-up of sharp gradients in the
Alfven wave front, in the transverse to its propagation 
direction.  
 It has been recently discovered \cite{2005A&A...435.1105T} in the solar coronal context that when the phase-mixing 
 mechanism is considered in the kinetic regime, electrons can be effectively accelerated by 
 the phase-mixed dispersive AW. It was found \cite{2005A&A...435.1105T} that when dispersive AW moves in the 
 transversely inhomogeneous plasma, parallel electric field is generated in the regions 
 of plasma inhomogeneity. This electric field can effectively accelerate electrons for a 
 wide range of plasma parameters \cite{2008PhPl...15k2902T}. It turns out that similar electron acceleration 
 mechanism has been also found in the Earth magnetosphere context \cite{1999JGR...10422649G,2004AnGeo..22.2081G}. 
 See also Ref.\cite{2006A&A...449..449M} for the cross-comparison. The difference is that  
 Ref. \cite{2005A&A...435.1105T,2008PhPl...15k2902T} have 
 studied the case of plasma over-density, transverse to the background magnetic field, 
 whereas Ref.\cite{1999JGR...10422649G,2004AnGeo..22.2081G,2006A&A...449..449M} 
 considered the case of plasma under-density (a cavity), as 
 dictated by the different applications considered (solar coronal loops and Earth 
 auroral plasma cavities). 
 Ref.\cite{2008GeoRL..3520101L} 
 studied the kinetic-scale Alfven wave phase mixing 
 at the boundaries of the density cavity, the strongly varying Alfven speed profile above the Earth 
 auroral ionosphere (so-called ionospheric Alfven resonator),
 which leads to the development of the parallel electric fields that accelerate electrons in the aurora.
 They performed simulations to study the evolution of these fields including both 
 parallel and perpendicular inhomogeneity. 
 Calculations of Ref.\cite{2003JGRA..108.8005L}, based on the linear kinetic theory of Alfven waves, also indicate that Landau
damping of these waves can efficiently convert the Poynting flux into field-aligned
acceleration of electrons.
  More recently, an analytical theory has been developed \cite{2011A&A...527A.130B} 
 to explain the numerical simulations presented in Ref.\cite{2005A&A...435.1105T,2008PhPl...15k2902T}. 
 Also, a possibility 
 of acceleration of electrons by inertial AWs has been explored in a one dimensional, 
 homogeneous plasma case (without phase mixing) \cite{2009ApJ...693.1494M}.
 The latter extended and adopted for solar coronal context
 earlier work of Ref.\cite{1994JGR....9911095K}
 who demonstrated the Fermi-like acceleration by 
 inertial Alfven waves of a small proportion of electrons, 
 up to speeds of twice the Alfven speed and argued that this is a plausible scenario 
 for auroral electron acceleration. 
 The effect of the Hall term on 
 the phase mixing of AWs has been studied in Ref.\cite{2011A&A...525A.155T}.

 This work aims to fill 
 the gap in understanding of particle acceleration by dispersive Alfven waves in the 
 transversely inhomogeneous plasma via full kinetic simulation particularly focusing on
 the effect of polarisation of the waves and different regimes (inertial of kinetic).
In particular, we study particle acceleration by the low frequency ($\omega=0.3\omega_{ci}$)
DAWs, similar to considered in Ref.\cite{2005A&A...435.1105T,2008PhPl...15k2902T},      
but here we focus on the effect of the wave polarisation, left- (L-) and right- (R-) circular and elliptical,
in the different regimes inertial ($\beta < m_e/m_i$) and kinetic 
($\beta > m_e/m_i$). The primary goal is to study the latter effects on the 
efficiency of particle acceleration and the parallel electric field generation.

\section{The model}

We use EPOCH (Extendible Open PIC Collaboration) a multi-dimensional, fully electromagnetic, 
relativistic particle-in-cell code which was developed and is used by 
Engineering and Physical Sciences Research Council (EPSRC)-funded 
Collaborative Computational Plasma Physics  (CCPP) consortium of 30 UK researchers.
We use 2.5D version 
of the EPOCH code which means that we have two spatial component along $x$- and $y$-axis 
and there are three $V_x,V_y,V_z$ particle velocity components
present (for electrons and ions).
The relativistic equations of motion are solved
for each individual plasma particle.
The code also solves Maxwell's equations, with self-consistent currents, using 
the full component set of EM fields
$E_x,E_y,E_z$  and $B_x,B_y,B_z$.
EPOCH uses un-normalised SI units, however, in order for our results
to be general, we use the normalisation for the graphical
presentation of the results as follows.
Distance and time are normalised to $c / \omega_{pe}$ and $\omega_{pe}^{-1}$, while
electric and magnetic fields to $\omega_{pe}c m_e /e$ and  $\omega_{pe} m_e /e$ respectively.
Note that when visualising the normalised results we use   $n_0 =10^{16}$ m$^{-3}$ in the
least dense parts of the domain ($y=0$ and $y=y_{max}$), which are located at the lowermost and uppermost edges of the
simulation domain
(i.e. fix $\omega_{pe} = 5.64 \times 10^9$ Hz radian on the domain edges).
Here $\omega_{pe} = \sqrt{n_e e^2/(\varepsilon_0 m_e)}$ is the electron plasma frequency,
$n_\alpha$ is the number density of species $\alpha$ and all other symbols have their usual meaning.
The spatial dimension of the simulation box are varied
as $x=5000$ and $y=200$ grid points for the mass ratio $m_i/m_e=16$, 
and $x=10712$ and $y=200$ grid points for the mass ratio $m_i/m_e=73.44$ which
represents 1/25th of the actual $m_i/m_e$ of 1836.
The choice of the mass ratio $m_i/m_e=1836/25=73.44$ is motivated
by following reasons: (i) since the DAWs considered
in this study are essentially Alfvenic (for the fixed throughout this paper 
value of $\omega =0.3\omega_{ci}$ there is very little difference
between Alfven  branch with $\omega=V_Ak$ and L- and R- circularly polarised
DAW branches that propagate along the magnetic field, see
e.g. Fig. 3.4 from Ref.\cite{dendy}), as the Alfven seed $V_A=B_0/\sqrt{\mu_0 m_i n}$
scales as $m_i^{-1/2}$ the actual Alfven wave would be 5 times slower than considered
here; (ii) $m_i/m_e=1836/25=73.44$ value lands us in the kinetic regime
because plasma beta in this study is fixed at 
$\beta= 2 (v_{th,i}/c)^2(\omega_{pi}/\omega_{ci})^2 = 
n_0(0,0)k_B T /(B_0^2/(2\mu_0))=0.02$. Thus $\beta=0.02 > m_e/m_i=1/73.44=0.0136$;
(iii) This is the maximal value that can be considered with the
available computational resources -- using 32  Dual Quad-core Xeon 
$= 32 \times 8 = 256$ processor cores. A typical with $m_i/m_e=73.44$ run 
takes 48 hours on 256 processors.
The grid unit size is  $\lambda_D$. 
Here $\lambda_D = v_{th,e}/ \omega_{pe}$
is the Debye length ($v_{th,e}=\sqrt{k_B T/m_e}$ is electron thermal speed).
We use 100 electrons and 100 ions per cell which means that for 
$m_i/m_e=16$ there are $2\times10^8$ and for $m_i/m_e=73.44$ there are 
$4.2848\times 10^8$ particles in the simulation box.
Plasma
number density $n(x,y)$ (and hence $\omega_{pe}(x,y)$) can be regarded 
as arbitrary, because we wish our results to stay general.
We impose constant background magnetic field $B_{0x} =320.753$ Gauss along
$x$-axis. This corresponds to $B_{0x}=1.0 (\omega_{pe} m_e /e)$, so that
with the normalisation used for the visualisation
purposes, normalised background magnetic field is unity. 
This fixes $\omega_{ce}/\omega_{pe} = 1.00$.
Electron and ion temperature at the simulation box edge
is also fixed at $T(0,0)=T_e(0,0)=T_i(0,0)=6\times10^7$K. 
This in conjunction with $n_0(0,0) =10^{16}$ m$^{-3}$
makes plasma parameters similar to that of a dense flaring loops
in the solar corona. For example radiative hydrodynamic 
modelling of the Bastille-Day solar flare by Ref.\cite{2004A&A...419.1149T} (see their Fig.1b and 1c)
indicate loop apex temperatures and densities as $5.3\times10^7$K 
and $2.5\times10^{17}$ m$^{-3}$ respectively.
This is consonant with our overall goal to provide
a viable, fully-kinetic model for particle acceleration by means
of DAWs during solar flares to alleviate so called "number-problem".
The latter refers to the too high total number ($10^{34}-10^{37}$ per second) 
of accelerated electrons required to produce the observed hard X-ray emission 
compared to that available in the corona, if the particle acceleration takes place
at the loop apex. This would mean that if the solar flare particle
acceleration volume
is in the range of $1-10$ Mm$^3$ with the number density of $n=10^{16}$ 
m$^{-3}$, to match the observational $10^{34}-10^{37}$ accelerated 
electrons per second, full 100\% of electrons need to be accelerated.
There is no known particle acceleration mechanism that operates
with the 100\% efficiency. Thus it would seem attractive
to have flare energy release triggering DAWs which rush
down the footpoints and accelerate electrons along the way to chromosphere
\cite{2008ApJ...675.1645F}. Such scenario is rather attractive in the light of
the earlier work Ref.\cite{2005A&A...435.1105T} which showed that such
particle acceleration is indeed possible.
The summary of numerical simulation parameters is given in
Table~\ref{pars}.

\begin{table}
\caption[]{Numerical simulation parameters}
\label{pars}
$$ 
\begin{array}{lll}
\hline
\noalign{\smallskip}
\mathrm{Regime} & \mathrm{Inertial} & \mathrm{Kinetic} \\
\noalign{\smallskip}
\hline
m_i/m_e      &  16 & 73.44\\
\omega_{ce}/\omega_{pe} & 1.000 & 1.000 \\
\beta & 0.020 & 0.020 \\
c/\omega_{pe} \mathrm{[m]} & 0.053 & 0.053 \\
\lambda_D=r_{L,e} \mathrm{[m]} & 0.005 & 0.005 \\
v_{th,e}/c & 0.101 & 0.101\\
v_{th,i}/c & 0.025 & 0.012\\
V_A/c=\omega_{ci}/\omega_{pi} & 0.25 & 0.117 \\
V_{A,ph}/c & 0.243 & 0.116 \\
V_L/c & 0.201 & 0.097 \\
V_R/c & 0.264 & 0.131 \\ 
t_{end}=75\omega_{ci}^{-1} \mathrm{[}\times10^{-7}\mathrm{s]} & 2.127 &9.763 \\
n_y & 200 & 200 \\
n_x & 5000 & 10712 \\
\noalign{\smallskip}
\hline
\end{array}
$$ 
\end{table}

\begin{figure*}[htbp]    
\centerline{\includegraphics[width=0.8\textwidth]{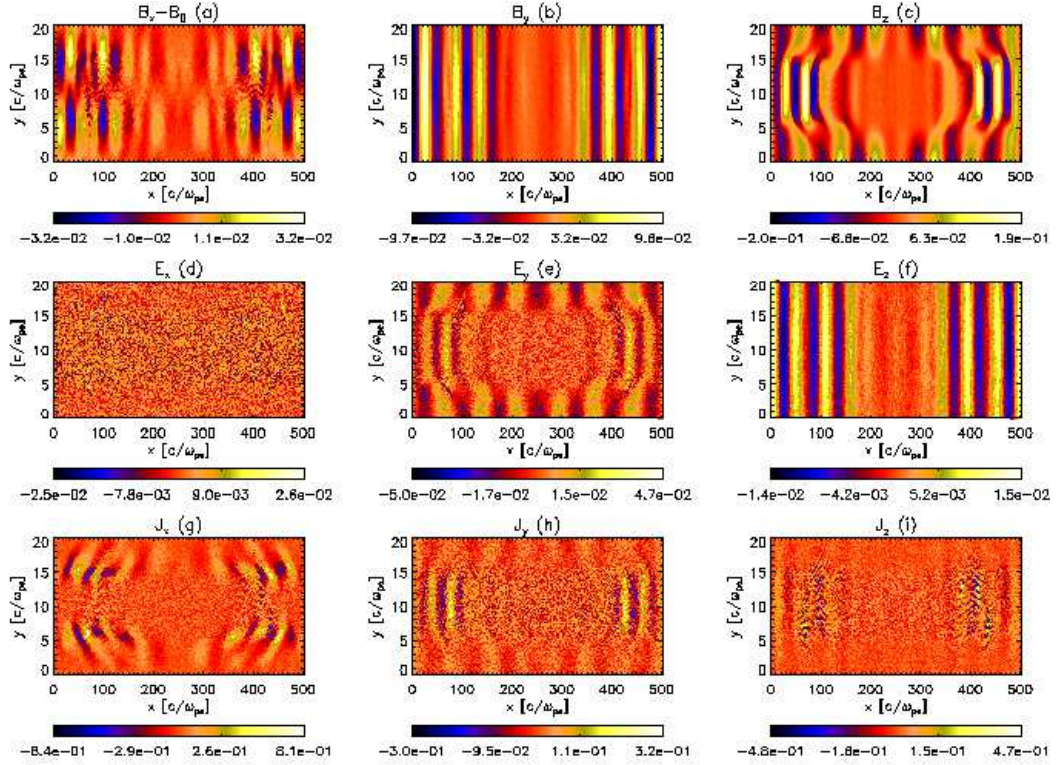}}
\caption{Contour (intensity) plots of the following physical quantities at $t_{end}=75\omega_{ci}^{-1}$ 
(the time when DAW develops three wavelengths): 
(a) $B_x(x,y,t=t_{end})-B_0$,
(b) $B_y(x,y,t=t_{end})$, (c) $B_z(x,y,t=t_{end})$,
(d) $E_x(x,y,t=t_{end})$, (e) $E_y(x,y,t=t_{end})$, (f) $E_z(x,y,t=t_{end})$,
(g) $J_x(x,y,t=t_{end})$, (h) $J_y(x,y,t=t_{end})$ and (i) $J_z(x,y,t=t_{end})$.
See text for the normalisation used.
This figure pertains to the numerical run L16 (see Table~\ref{runs} for the run parameters).
}
\end{figure*}

As in Ref.\cite{2005A&A...435.1105T}, in order to model solar coronal 
loops, we consider a transverse to the background magnetic field 
variation
of number density as following
\begin{equation}
 {n_i(y)}=
{n_e(y)}=1+3 \exp\left[-\left(\frac{y-100\lambda_D}{50 \lambda_D}\right)^6\right]
\equiv f(y).
\end{equation}
Eq.(1) implies that in the central region (across the 
$y$-direction), the density is
smoothly enhanced by a factor of 4, and there are the 
strongest density gradients having 
a width of about ${30 \lambda_D}$ around the 
points $y=51.5 \lambda_D$ and $y=148.5 \lambda_D$.
The background temperature of ions and electrons, 
are varied accordingly
\begin{equation}
{T_i(y)}/{T_0}=
{T_e(y)}/{T_0}=f(y)^{-1},
\end{equation}
such that the thermal pressure remains constant. Because the background magnetic field
along the $x$-coordinate  is constant, the total pressure is constant too.
Thus the pressure balance in the initial conditions
is fully enforced.
Further we drive domain left edge, $x=1\lambda_D$, as follows
\begin{eqnarray}
E_y(1,y,t+\Delta t)= E_y(1,y,t)- &   \nonumber \\
A_y\sin(\omega_d t)\left(1-\exp\left[-(t/t_0)^2\right]\right),& 
\end{eqnarray}
\begin{eqnarray}
E_z(1,y,t+\Delta t)=E_z(1,y,t)- &   \nonumber \\
A_z\cos(\omega_d t)\left(1-\exp\left[-(t/t_0)^2\right]\right),&
\end{eqnarray}
for the case of left-polarised DAW and 
\begin{eqnarray}
E_y(1,y,t+\Delta t)=E_y(1,y,t)+ &   \nonumber \\
A_y\sin(\omega_d t)\left(1-\exp\left[-(t/t_0)^2\right]\right),&
\end{eqnarray}
\begin{eqnarray}
E_z(1,y,t+\Delta t)=E_z(1,y,t)- &   \nonumber \\
A_z\cos(\omega_d t)\left(1-\exp\left[-(t/t_0)^2\right]\right),&
\end{eqnarray}
for the case of right-polarised DAW.
Here $\omega_d$ is the driving frequency which was fixed at $\omega_d=0.3\omega_{ci}$
(Although numerical value of $\omega_d$ actually varies when ion mass is varied).
This ensures that no significant ion-cyclotron damping is present and also
that the generated DAW is well on the Alfven wave branch with dispersion
properties similar to the Alfven wave (see e.g. Fig. 3.4 from Ref.\cite{dendy}). 
$t_0$ is the onset time of the driver, which was fixed at $3.0 \omega_{ci}^{-1}$
i.e.  $48.000  \omega_{pe}^{-1}$ for the case of $m_i/m_e=16$ and
$220.320  \omega_{pe}^{-1}$ for the case of $m_i/m_e=73.44$. 
This means that the driver onset time is about $3^2=9 \omega_{ci}^{-1}$. 
Imposing such a transverse electric field $E_\perp=E_{y,z}$ driving 
on the system results in the generation of
L- or R- circularly or elliptically polarised DAW.
Definition for the L- or R- polarisation is a follows:
Consider the magnetic field directed along positive $x$-axis direction 
and electric field vector $\vec E$
rotates in $yz$-plane. 
In the case of L-polarisation,
if left hand thumb points along the magnetic field (and hence
along the propagation direction of DAW) curled fingers show the direction of
rotation of $\vec E$, i.e. clockwise. In the case of R-polarisation 
then right hand thumb points along the magnetic field 
and curled fingers show the direction of
rotation of $\vec E$, i.e. counterclockwise.
Note that ions rotate in the same direction as  $\vec E$ of the L-polarised
DAW (that is why ion-cyclotron resonance can occur), 
at the same time electrons rotate in the opposite sense.
For the DAW the dispersion relation is  as following (see Eqs.(3.78) and (3.87) from Ref.\cite{dendy}):
\begin{equation}
k=\frac{\omega}{c}\left(1+\frac{\omega_{pe}^2+\omega_{pi}^2}{(\omega_{ce}\pm \omega)
(\omega_{ci} \mp \omega)}\right)^{1/2}
\end{equation}
where upper signs are for the L-polarisation and lower signs for the R-polarisation.
Eq.(7) shows that for L-polarisation the allowed frequency range of $0<\omega<\omega_{ci}$
which indicates existence in the ion-cyclotron resonance at $\omega=\omega_{ci}$.
Ions can resonate because they rotate in the same direction as
the electric field vector of the DAW.
For the R-polarisation the allowed range is $0<\omega<\omega_{ce}$ with the electron-cyclotron 
resonance at $\omega=\omega_{ce}$. In the plasma literature L-polarised wave is usually
referred to as ion cyclotron wave because it cuts off at $\omega=\omega_{ci}$,
while R-polarised wave is referred to as the whistler branch as it cuts off at $\omega=\omega_{ce}$.
We follow the terminology adopted in \cite{2000SSRv...92..423S} and refer to
all waves with perpendicular wavelengths comparable to any of the above
mentioned kinetic scales as DAWs, irrespective whether they are on L- or R-polarisation
dispersion branch and inertial or kinetic regime.
For $\omega \leq 0.3 \omega_{ci}$  from the dispersive properties point of view
there is little difference between L- and R-polarised waves and the shear Alfven wave
which has the phase speed of $V_{A,ph}/c=(V_A/c)/\sqrt{1+(V_A/c)^2}$ which
for the parameters considered is slightly different from the Alfven speed
$V_A/c=\omega_{ci}/\omega_{pi}$. Eq.(7) has been used to calculate 
L- and R- polarisation wave phase seeds in Table~\ref{pars}.

\begin{figure}[htbp]    
\centerline{\includegraphics[width=0.49\textwidth]{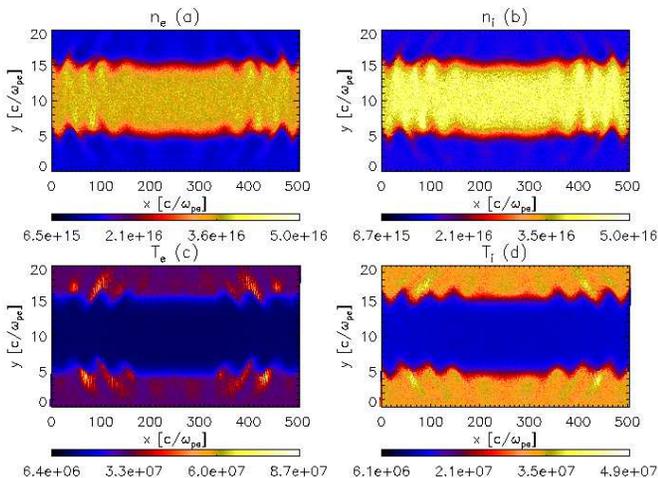}}
\caption{Contour (intensity) plots of the following physical quantities at $t_{end}=75\omega_{ci}^{-1}$ 
(the time when DAW develops three wavelengths): 
(a) $n_e(x,y,t=t_{end})$, (b) $n_i(x,y,t=t_{end})$, (c) $T_e(x,y,t=t_{end})$, and (d) $T_i(x,y,t=t_{end})$.
Here physical quantities are in SI units.
This figure pertains to the numerical run L16.}
   \end{figure}

The initial amplitudes of the $E_\perp$ are chosen 
as following: (i) in the case of circular polarisation
$A_y=A_z=0.05(cB_{x0})$; Thus,  
the relative DAW amplitude is 5\% of the background,
thus the simulation is weakly non-linear.
(ii) in the case of elliptical polarisation when electric field in the ignorable $z$
direction is six times larger than in $y$ direction, 
$A_y=0.05(cB_{x0})\times\sqrt{2/37}$ and $A_z=0.05(cB_{x0})\times\sqrt{2/37}\times 6$;
(iii) in the case of elliptical polarisation when electric field in the ignorable $z$
direction is six times smaller than in $y$ direction,
$A_y=0.05(cB_{x0})\times\sqrt{2/37}\times 6$ and $A_z=0.05(cB_{x0})\times\sqrt{2/37}$.
Note that the amplitudes $A_y$ and $A_z$ have been chosen such that 
{\it in all considered cases of circular or elliptical polarisation,
the wave power pumped into the system is the same}; i.e. in all presented runs
$A_y^2+A_z^2=const$, with the constant being the same for each numerical run.
This ensures consistency when studying the efficiency of particle acceleration
and $E_\parallel$ generation.

\begin{figure*}[htbp]    
\centerline{\includegraphics[width=0.65\textwidth]{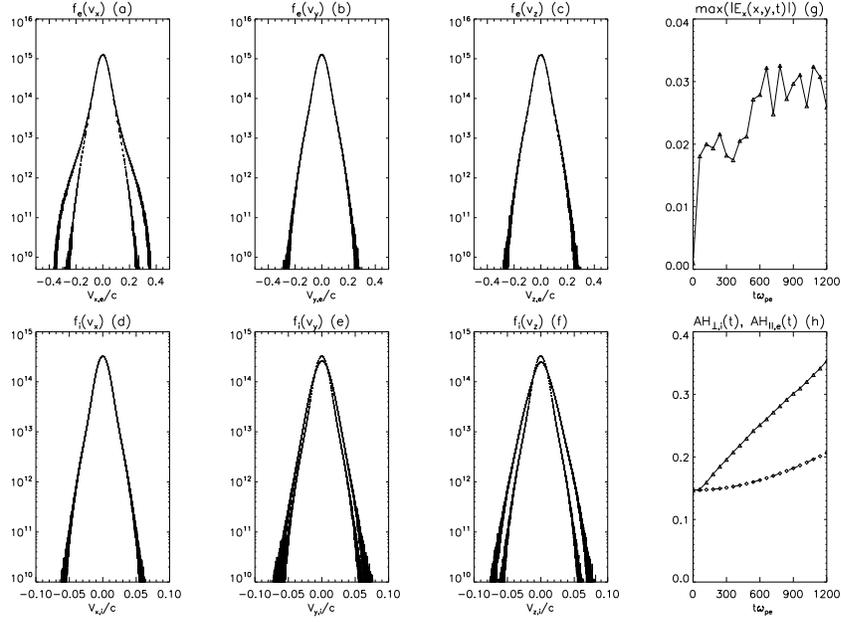}}
\caption{Time evolution (at $t=0$ and $t=t_{end}=75\omega_{ci}^{-1}$) of electron and ion velocity distribution 
functions (on a log-linear plot): 
(a) $f_e(v_x,t=0)$ dashed (inner) curve and $f_e(v_x,t=t_{end})$ solid (outer) curve,
(b) $f_e(v_y,t=0)$ dashed (inner) curve and $f_e(v_y,t=t_{end})$ solid (outer) curve (note that the both curves overlap),
(c) $f_e(v_z,t=0)$ dashed (inner) curve and $f_e(v_z,t=t_{end})$ solid (outer) curve (note that the both curves overlap), 
(d) $f_i(v_x,t=0)$ dashed (inner) curve and $f_i(v_x,t=t_{end})$ solid (outer) curve (note that the both curves overlap),
(e) $f_i(v_y,t=0)$ dashed (inner) curve and $f_i(v_y,t=t_{end})$ solid (outer) curve,
(f) $f_i(v_z,t=0)$ dashed (inner) curve and $f_i(v_z,t=t_{end})$ solid (outer) curve.
Time evolution (at 20 time intervals between $t=0$ and $t=t_{end}$) of the
following: (g) $\max |E_x(x,y,t)|$, triangles connected with a solid curve and
(h) $AH_{\parallel,e}(t)$ index, diamonds connected with dashed curve, according to
Eq.(8), 
$AH_{\perp,i}(t)$ index, triangles connected with a solid curve, according to
Eq.(9).
This figure pertains to the numerical run L16.}
   \end{figure*}

The considered numerical runs with their identifying name used throughout this paper is shown in
Table~\ref{runs}.

\begin{table}
\caption[]{Numerical simulation run identification and physical parameters. I stands for inertial
and K for kinetic. Reg. stands for regime.}
\label{runs}
$$ 
\begin{array}{lllllll}
\hline
\noalign{\smallskip}
\mathrm{Run}\;\mathrm{ID} &  \mathrm{Polaris.} & m_i/m_e & \mathrm{Reg.} & A_y/A_z & t_{end} [\omega_{ci}^{-1}]&\mathrm{Figs.}\\
\noalign{\smallskip}
\hline
\mathrm{L16} & \mathrm{L-circ.} & 16 & \mathrm{I} & 1& 75 & 1,2,3\\
\mathrm{R16} & \mathrm{R-circ.} & 16 & \mathrm{I} & 1& 75 & 4\\
\mathrm{EL16} & \mathrm{L-ellip.} & 16 & \mathrm{I} & 6& 75 & 5\\
\mathrm{ER16} & \mathrm{R-ellip.} & 16 & \mathrm{I} & 6& 75 & 6\\
\mathrm{EL16}_1 & \mathrm{L-ellip.} & 16 & \mathrm{I} & 1/6& 75 & 7\\
\mathrm{ER16}_1 & \mathrm{R-ellip.} & 16 & \mathrm{I} & 1/6& 75 & 8\\
\mathrm{L73} & \mathrm{L-circ.} & 73.44 & \mathrm{K} & 1& 75 & 9\\
\mathrm{R73} & \mathrm{R-circ.} & 73.44 & \mathrm{K} & 1& 75 & 10,11,12\\
\mathrm{EL73} & \mathrm{L-ellip.} & 73.44 & \mathrm{K} & 6& 75 & 13\\
\mathrm{ER73} & \mathrm{R-ellip.} & 73.44 & \mathrm{K} & 6& 75 & 14\\
\mathrm{EL73}_1 & \mathrm{L-ellip.} & 73.44 & \mathrm{K} & 1/6& 75 & 15\\
\mathrm{ER73}_1 & \mathrm{R-ellip.} & 73.44 & \mathrm{K} & 1/6& 75 & 16\\
\mathrm{L16Long} & \mathrm{L-circ.} & 16 & \mathrm{I} & 1& 300 & 17,18\\
\noalign{\smallskip}
\hline
\end{array}
$$ 
\end{table}

\section{Results}

Below we present numerical simulation results for the runs given in Table~\ref{runs}.
Run L16 is similar to one considered in \cite{2005A&A...435.1105T}.
The latter used numerical code TRISTAN. Throughout this paper we use numerical code EPOCH.
The run L16 provides important benchmarking and validation for our results. 
We gather from Fig.~1(e) and 1(f) that transverse electric field driving, as described above,
generates DAWs that for $0<x<250$ propagate to the right, i.e. along the magnetic field.
However, due to the driving at $x=1$ also generates a wave that propagates to the left, because of the
periodic boundary conditions used both for particles and fields throughout this paper, the
left-propagating wave appears on the right hand edge of the domain and propagates in the domain
$250 <x <500$. In turn, such driving also generates transverse magnetic fields which are
shown in Figs.~1(b) and 1(c). One can clearly see from Figs.~1(c) and 1(e) that
DAW components $(E_y,B_z)$ experience phase mixing. That is initially flat wave front
becomes progressively distorted, because in the middle of the domain phase speed of the wave is
slower than at the edges. 
Figs.~1(b) and 1(c) show that DAW has developed three full wavelength. Period of the
DAW is $T=2\pi/\omega_d= 2\pi/(0.3\omega_{ci})\approx 21\omega_{ci}^{-1}$.
Three periods (and hence three wavelengths) will take $63\omega_{ci}^{-1}$ to develop 
which added to the DAW driver on-set time of $9\omega_{ci}^{-1}$ gives $72\omega_{ci}^{-1}$
which roughly coincides with the end of the simulation time of $75\omega_{ci}^{-1}$ (shortfall of
$75-72=3\omega_{ci}^{-1}$ could be roughly attributed for the latency time
it takes for ions to respond to DAW driving).
\begin{figure*}[htbp]    
\centerline{\includegraphics[width=0.65\textwidth]{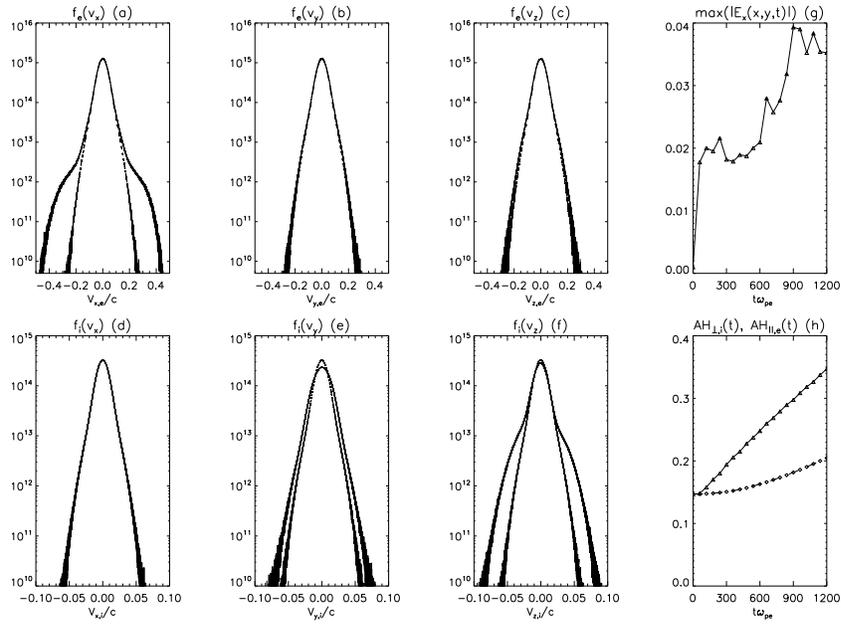}}
\caption{As in Fig.~3 but for the numerical run R16.}
   \end{figure*}

We gather from Figs.1(a) and 1(d) that in the regions where density has
 transverse gradients longitudinal (parallel) magnetic field ($B_x$) and electric field ($E_\parallel=E_x$) 
are generated. It can be clearly seen that these are generated only on the density gradients
located within $4 <y <7$ and $13 < y < 16$. Outside these ranges as can be corroborated
by plotting $n(y)$ from Eq.(1) the plasma density is constant.
This means the transverse gradient scale is $7-4=16-13=3 c /\omega_{pe}\approx 30 \lambda_D$
for $m_i/m_e=16$ corresponds to $0.75 c/\omega_{pi}$, i.e. the transverse density
gradient is about one ion-inertial lengths long.
Further insight can be obtained by looking at $x,y,z$ components
of the plasma current $\vec J = e (n_i \vec V_i -  n_e \vec V_e)$.
Our normalisation for the current is $J_{norm}=e \langle n_e \rangle \langle v_{th,e} \rangle$
where $\langle n_e \rangle$ is the number density averaged over the transverse, $y$-direction
and $\langle v_{th,e} \rangle =\sqrt{k_B \langle T_e \rangle /m_e}$ with $\langle T_e \rangle$ being electron
temperature  also averaged over $y$-direction.
In effect $J_{norm}$ is a typical electron thermal velocity current.
We gather from Fig.~1(g) that sizable longitudinal currents $(\approx 0.8)$
are generated in the transverse density gradient regions.
These currents provide  explanation for the source of the
 $E_\parallel$. As also outlined by Ref.\cite{2007NJPh....9..262T},
 charge separation between more mobile electrons and less mobile ions
 causes generation of time variable $E_\parallel$.
 The transverse currents $J_y$ (Fig.~1(h)) and $J_z$ (Fig.~1(i))
 are manifestations of the driving DAW which actually
follow the $E_y$ and $E_z$ spatial pattern respectively.

Fig.~2 displays behaviour of number density and temperature
of electrons and ions. The noteworthy features are:
(i) the
transverse (kink-like) oscillations of plasma density that are mainly
confined the density gradient regions (Figs.~2(a) and 2(b));
(ii) hot plumes of plasma $\approx 8.7 \times 10^7$K
that emanate from the density (and temperature) 
gradient regions where particle acceleration by DAWs takes place
(Figs.~2(c) and 2(d)).
Fig.2 also gives a good indication how the initial
pressure balance looks like: number density goes up from
$10^{16}$ cm$^{-3}$ on the domain top ($y=L_{y,max}=20.12$) and bottom ($y=0$) edges and
increases to $4\times10^{16}$ cm$^{-3}$ in the middle of the $y$-range.
   
\begin{figure*}[htbp]   
\centerline{\includegraphics[width=0.65\textwidth]{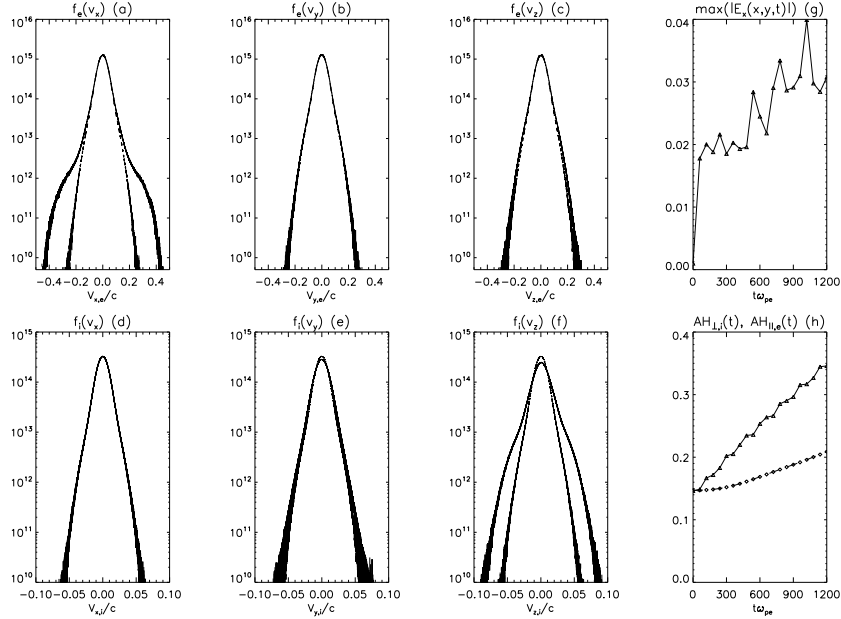}}
\caption{As in Fig.~3 but for the numerical run EL16.}
   \end{figure*}
 
Fig.3 shows the distribution function time evolution, 
quantifies the generated parallel electric field  amplitude
and associated particle acceleration for the run L16.
We gather from Figs.~3(a)-3(f) that as in 
\cite{2005A&A...435.1105T}
the generated parallel electric field in the
density in homogeneity regions produces electron beams
propagating along the magnetic field (Fig.~3(a)), while
in the transverse directions $y$ and $z$ the electron distribution
function remains unchanged (Fig.~3(b) and 3(c)).
In the case of ions only temperature increase is
seen in the transverse direction to the magnetic field.
Note that Ref.\cite{2011PhPl...18c0702D} also found heating of ions with L-circularly
polarised AWs using test-particle simulations in the context of
solar chromospheric heating.
It should be noted that Ref.\cite{2011PhPl...18c0702D} considers non-resonant
wave-particle interactions, in which the polarization, as such, 
is not important.
No ion beam have developed as we see only broadening of the 
distribution function (Fig.~3(e) and 3(f)).
Along the magnetic field the ion distribution function remains unchanged (Fig.3(d)).
The evolution of the amplitude of the parallel electric field
is studied in Fig.3(g) where we plot $\max|E_x(x,y,t)|$.
Since parallel electric field is mostly confined to the
density gradient regions $\max|E_x(x,y,t)|$ as function of time
essentially tracks the amplitude of the generated parallel electric field
in the density gradients. We see that after initial growth it saturates at about
0.03. We quantify the  particle acceleration and plasma heating 
by defining the following indexes:
\begin{equation}
AH_{\parallel,e}(t)=
\frac{\int_{|v_{x}| > \langle v_{th,e}\rangle}^\infty      f_e(v_x,t)dv_x\Biggl/\left(2 L_{IH,y}\times L_{x,max}\right)}
 { \int_{-\infty}^\infty   f_e(v_x,0) dv_x \Biggl/\left( L_{y,max}\times L_{x,max}\right)},
\end{equation}
\begin{equation}
AH_{\perp,i}(t)=
\frac{\int_{|v_{\perp}| > \langle v_{th,i}\rangle}^\infty      f_i(v_\perp,t)dv_\perp\Biggl/\left(2 L_{IH,y}\times L_{x,max}\right)}
 { \int_{-\infty}^\infty   f_i(v_\perp,0) dv_\perp \Biggl/\left( L_{y,max}\times L_{x,max}\right)}.
\end{equation}

\begin{figure*}[htbp]    
\centerline{\includegraphics[width=0.65\textwidth]{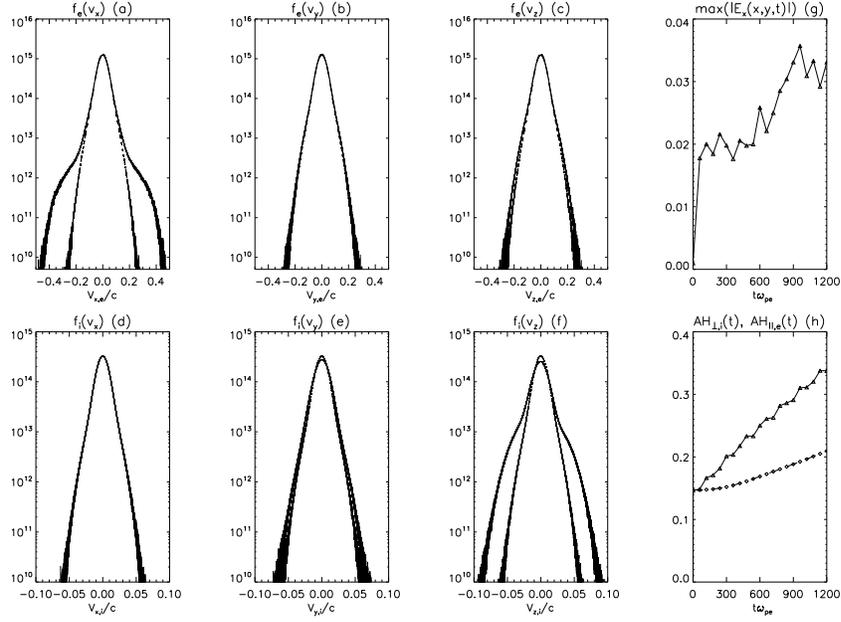}}
\caption{As in Fig.~3 but for the numerical run ER16.}
   \end{figure*} 
   
Here $f_{e,i}$ are electron or ion velocity distribution functions. They are normalised in such a way that when integrated by all spatial and velocity
components they give the total number of real electrons and ions (not pseudo-particles that usually mimic much larger number of real particles
in the Particle-In-Cell simulation). 
$L_{IH,y}$ is the width of each density gradients according to Eq.(1), located within $4 <y <7$ and $13 < y < 16$.
$L_{y,max}$ is the full width in the $y$-direction. The 
numerator of Eq.(8) gives the number of superthermal electrons in the parallel to the
magnetic field direction, 
as a function of time,  divided by the area ($2 L_{IH,y}\times L_{x,max}$) where the electron acceleration takes place. 
As we know from Fig.~1
parallel electric field is generated only in the two density gradient regions. 
The denominator gives total number of 
real electrons in the whole simulation box, at $t=0$, divided by the total area ($L_{y,max}\times L_{x,max}$). Thus, the definition 
used in Eq.(8) effectively provides the fraction
(the percentage) of accelerated electrons in the density gradient regions.
The same logic applies to ions as reflected by Eq.(9) but in the transverse direction.
The choice for such indexes has several reasons: 
(i) $AH_{\parallel,e}(t)$ and $AH_{\perp,i}(t)$ quantify fraction of electrons and ions at time $t$ with velocities greater than average thermal 
velocity in density gradient regions, in the parallel and perpendicular direction respectively;
(ii) As can be seen from Figs.3-9 and 12-16,
driving by DAWs mostly affect electron motion parallel to the magnetic field, while for ions the effect is only in the perpendicular direction;
(iii) They way how $AH_{\parallel,e}(t)$ and $AH_{\perp,i}(t)$ are defined alleviates the problem that the distribution function value varies
as the total number of particles in the simulation is altered in different runs (because domain size needs to be adjusted so that
DAWs do not collide producing interference). 
For example comparing Fig.~3(a) and 9(a) we see that the height of the peak at $v_{x,e}=0$ is different despite the fact that both graphs
show the distribution function for the medium with the same temperature distribution and other parameters.
This is due to the fact that distribution function is normalised such that when integrated by all spatial and velocity components,
it gives the total number of particles (not pseudo-particles) in the simulation. Fig.~9 pertains to the run L73 in which number of
grids ($n_x=10712$) is more than twice compared to the run L16 in which $n_x=5000$. Since we always load 100 electrons and 100 ions
per cell, this results in doubling the peak height in Fig.9(a) compared to Fig.~3(a), as there are twice as many particles
in run L73 than in run L16. It is clear from the definition (Eq.(8))
that this difference in heights cancels out, thus providing the true precentage of accelerated electrons in the density gradient regions.
The only drawback our  $AH_{\parallel,e}(t)$ and $AH_{\perp,i}(t)$ indexes have is that they cannot distinguish between particle acceleration
(in which case bumps on the distribution function appear) and heating (in which case distribution function simple broadens, 
proportionately retaining its shape at $t=0$). Despite this $AH_{\parallel,e}(t)$ and $AH_{\perp,i}(t)$ indexes 
still provide useful information about the fraction of particles at time $t$ with velocities greater than average thermal 
velocity in the density gradient regions. 
We gather from Fig.~3(h) that $AH_{\parallel,e}(t)$ and $AH_{\perp,i}(t)$ indexes attain values 0.2 and 0.35, respectively.
This means that in the density gradient regions 20\% of electrons are accelerated in the parallel to the
magnetic field direction, and 35\% of ions gained velocity above thermal (were heated up) in the transverse direction.
It should be noted that these conclusions apply for the end simulation time of $t_{end}=75\omega_{ci}^{-1}$.
See discussion of run L16Long which discusses long term dynamics that
corresponds to $t_{end}=300\omega_{ci}^{-1}$.

\begin{figure*}[htbp]    
\centerline{\includegraphics[width=0.65\textwidth]{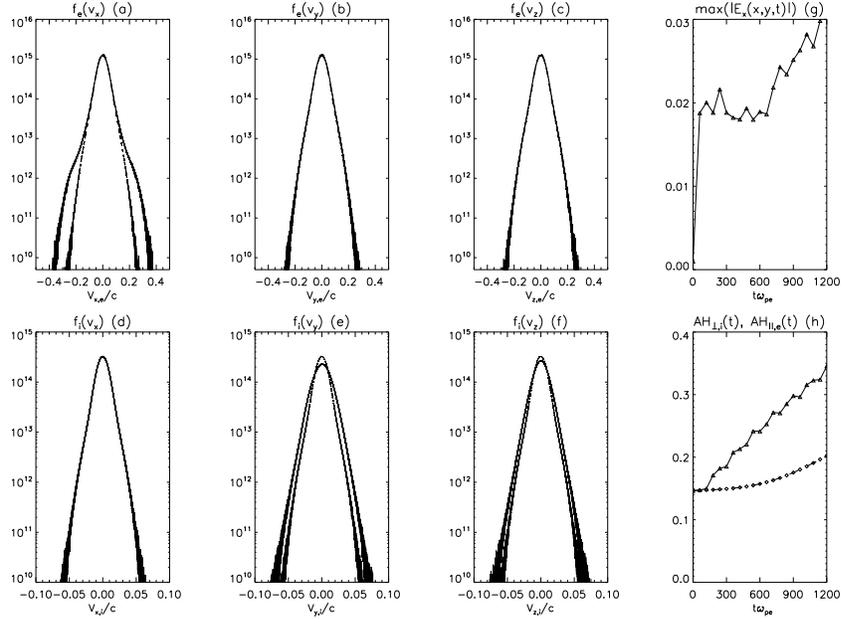}}
\caption{As in Fig.~3 but for the numerical run EL16$_1$.}
   \end{figure*}

It is interesting to note that in Ref.\cite{2008PhPl...15k2902T} we used
much more computationally demanding definition (see Eq.(5) from Ref.\cite{2008PhPl...15k2902T}).
The demanding part is storing large data for the distribution function ($f(v_{x,e},x,y)$) in order
to perform integration by $dxdy$. Our above definition (Eq.(8)) is much less demanding because
it only stores $f(v_{x,e})$ and spatial integration is replaced by the factor $1/(2 L_{IH,y} / L_{y,max})$.
By comparing the rightmost data point for  $AH_{\parallel,e}(t)$ in Fig.~3(h) with middle data point
from Fig.~2(b) from Ref.\cite{2008PhPl...15k2902T} (because their physical parameters are the same), we gather that they both yield value of 0.2,
thus proving that the both definitions are quivalent.
Another observation we make is that $AH_{\parallel,e}(t)$ starts from about 0.15.
Which is reasonable because it is known that for a Maxwellian distribution (which holds before the particle acceleration takes place)
16\% of particles have velocty above thermal. This, of course, applies to a Maxwellian
with a uniform temperature. In our case plasma temperature varies across $y$ coordinate, 
thus the Maxwellian does not appear as 
a parabola in e.g. Fig.~3(a), as is it should have done because it is a log-linear plot.

\begin{figure*}[htbp]   
\centerline{\includegraphics[width=0.65\textwidth]{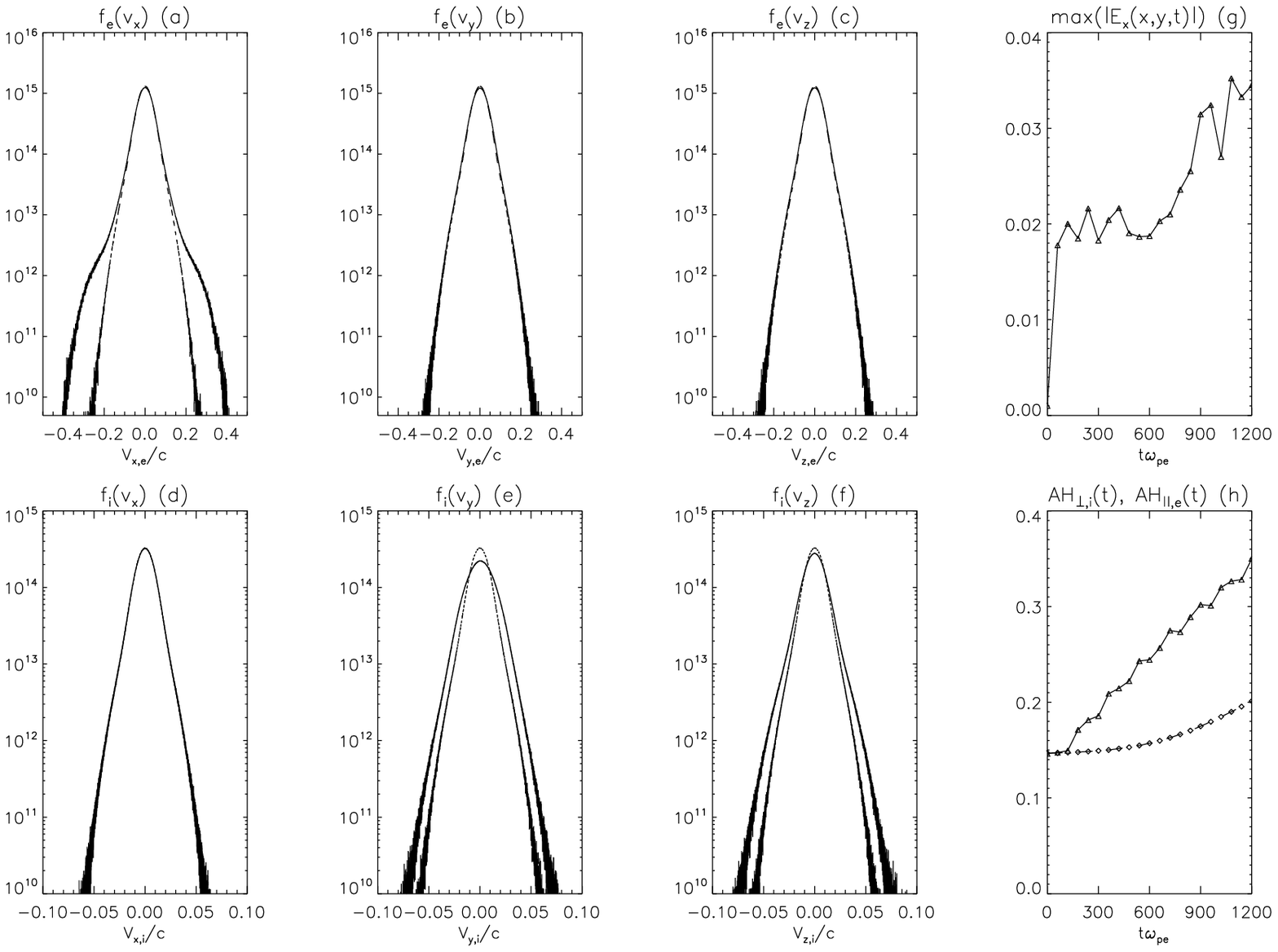}}
\caption{As in Fig.~3 but for the numerical run ER16$_1$.}
   \end{figure*}

Figs.~4, 5 and 6 correspond to the runs R16, EL16 and ER16. We group them together because their behaviour is similar,
but notably different from run L16 (Fig.~3). 
We see from Figs.~4(a), 5(a) and 6(a) that bumps on the electron parallel velocity distribution function
are much more pronounced and they now extend above 0.4c. 

In the case of ions
we gather from Figs.~4(f), 5(f) and 6(f) that transverse beams start to form in the ignorable $z$ spatial direction.
As far as the generated parallel electric field, in the density gradient regions, it saturates at about the same level of
0.03. $AH_{\parallel,e}(t)$ and $AH_{\perp,i}(t)$ indexes also exhibit similar behaviour. This is due to the fact that more
efficient electron and ion acceleration produces relatively small number of more energetic particles, which
do not affect in the indexes greatly. 

\begin{figure*}[htbp]    
\centerline{\includegraphics[width=0.65\textwidth]{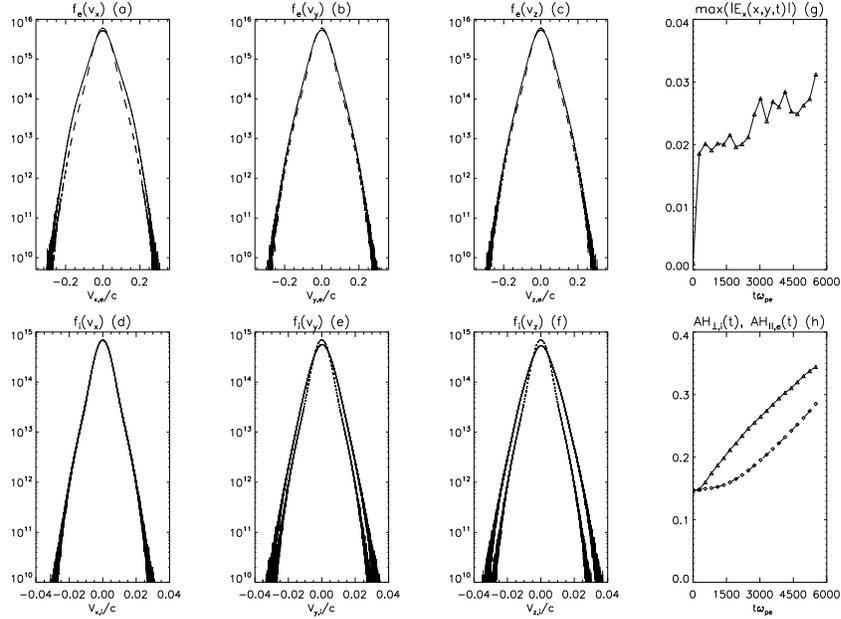}}
\caption{As in Fig.~3 but for the numerical run L73.}
   \end{figure*}

Intermediate conclusions that can be drawn from comparing runs L16, R16, EL16 and ER16 are that 
in the inertial  regime, R-circularly polarised, L-and R-elliptically polarised DAW with 
electric field in  the non-ignorable ($y$) transverse direction exceeding 6 times that of
in the ignorable ($z$) direction produce more pronounced parallel electron beams (in $v_x$) and transverse ion beams
in the ignorable ($v_z$) direction.
The electron beam speed coincides with the DAW seed.
We gather from Fig.~3(a) that accelerated beam electrons have speeds $0.25c<|v_{x,e}|<0.35c$
and from Figs.~4(a), 5(a) and 6(a) we have $0.25c<|v_{x,e}|<0.45c$.
We note from Table~\ref{pars} that phase speed of DAWs is $0.2c-0.26c$.
Thus electron acceleration seems to be due to the wave-particle interaction effects.
Namely, due to the Landau damping in which wave gives up energy whilst
accelerating particles. Particles which are faster than the wave give up their
energy to the wave, while particles which are slower gain energy from the wave.
For a given phase speed of the wave, $V_{A,ph}$, in the Maxwellian distribution there are always
more particles with  $v<V_{A,ph}$ than particles with 
$v>V_{A,ph}$. Thus, the net result of having Maxwellian distribution is that the wave gives up its momentum to the particles
with $v\leq V_{A,ph}$ and accelerates them. As a result the accelerated particle number increases around 
$v\approx V_{A,ph}$, i.e. a beam forms with $v\approx V_{A,ph}$.
It is interesting to note that 
similar conclusion has been drawn from theoretical calculations \cite{2011A&A...527A.130B}.
It should be stressed that their calculations were performed in the {\it kinetic}
regime. Whereas the above conclusions are in the {\it inertial} regime (because of the artificially lowered mass ratio).
We also note that similar conclusions still hold in the kinetic regime 
in our simulations (see below). Thus electron acceleration can be understood
by simple Landau damping of DAWs despite of the regime (inertial or kinetic).
Importance of Landau damping in this context was also pointed out by Ref.\cite{2003JGRA..108.8005L}.
   
\begin{figure*}[htbp]   
\centerline{\includegraphics[width=0.8\textwidth]{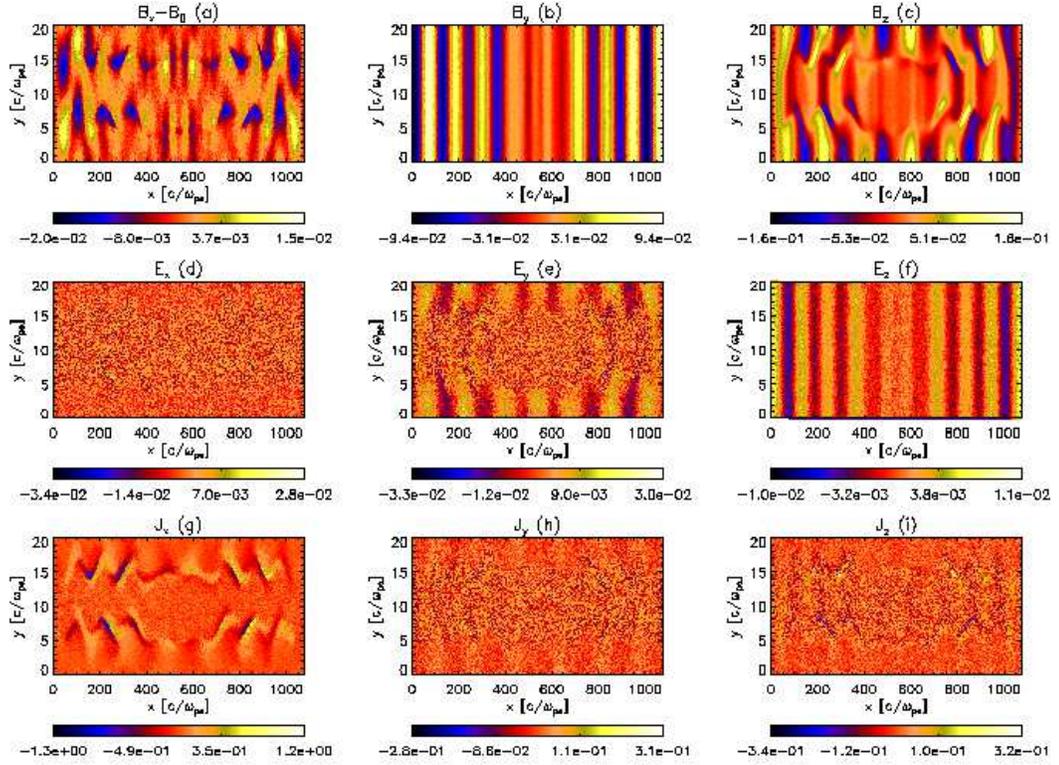}}
\caption{As in Fig.~1 but for the numerical run R73.}
   \end{figure*}

Figs.~7 and 8 correspond to runs EL16$_1$ and  ER16$_1$.
We see from these figures that the behaviour is similar to run L16.  
L-and R-elliptically polarised DAW with 
electric field in  the non-ignorable ($y$) transverse direction being $1/6$th that of
in the ignorable ($z$) direction produce weak parallel electron beams and weak
transverse ion heating (no pronounced ion beams).
   
\begin{figure}[htbp]    
\centerline{\includegraphics[width=0.49\textwidth]{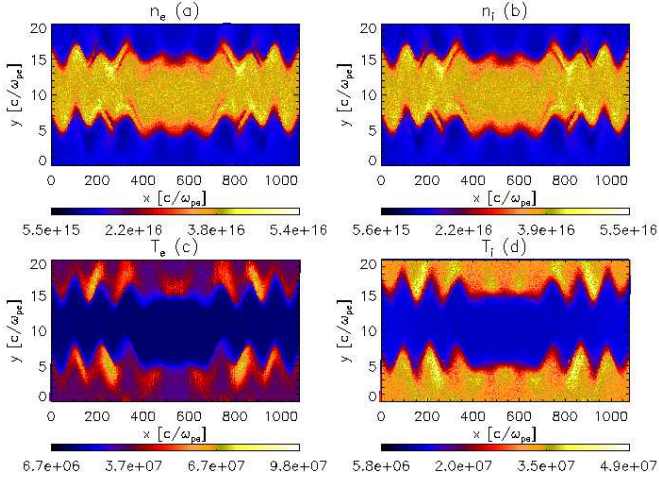}}
\caption{As in Fig.~2 but for the numerical run R73.}
   \end{figure}
   
Fig.~9 shows the results for the run L73 which brings us into the kinetic regime.
By comparing this run to L16 which is in the inertial regime we note following
differences: (i) bumps in the parallel electron velocity distribution function,
which represent the electron beam created by the parallel electric field,
now shift to the lower velocities which correspond to the phase speed
in the DAW. Table~\ref{pars} provides DAW phase speed of $\approx 0.1c$,
which indeed corresponds to the bump in the $f(v_{x,e})$ (in Fig.~9(a)) at $t=75 \omega_{ci}^{-1}$.
(ii) $AH_{\parallel,e}(t)$ index now attains 29\% which is about 50\% larger
than that of the run L16. This can be explained by the fact that at lower
velocities there more particles with lower velocity.
At present $m_i/m_e=1836$ is not accessible due to high computational costs.
But from run L73 it is clear that more massive ions would yield even higher
fraction of accelerated electrons with speeds comparable to the DAW
speed.

\begin{figure*}[htbp]   
\centerline{\includegraphics[width=0.65\textwidth]{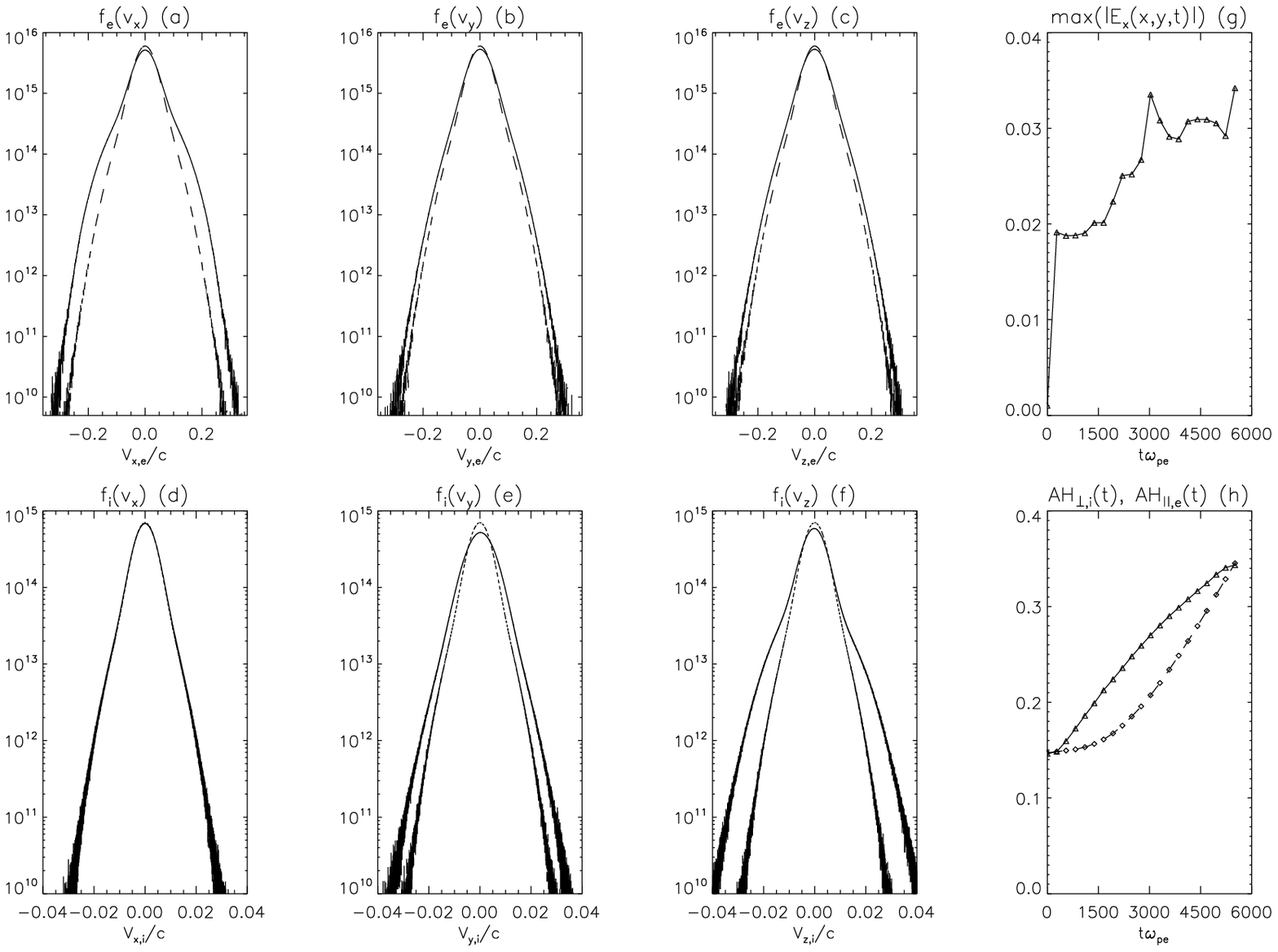}}
\caption{As in Fig.~3 but for the numerical run R73.}
   \end{figure*}       

Fig.~10 is analogous to Fig.~1 but here mass ratio is 73.44, as it corresponds to the
run R73. Note that as we have increased ion mass from 16 to 73.44 we had to about double
the domain length in order to prevent the two DAWs from colliding with each other,
so the results are not affected by the wave interference. As can be seen
from Fig.~10 phase mixing of DAWs develops in the kinetic regime
in a similar way as in the the inertial regime. We note that in the kinetic regime
parallel currents are stronger 1.3 (Fig.~10(g)) compared
to the inertial regime 0.83 (Fig.~1(g)).
   
\begin{figure*}[htbp]    
\centerline{\includegraphics[width=0.65\textwidth]{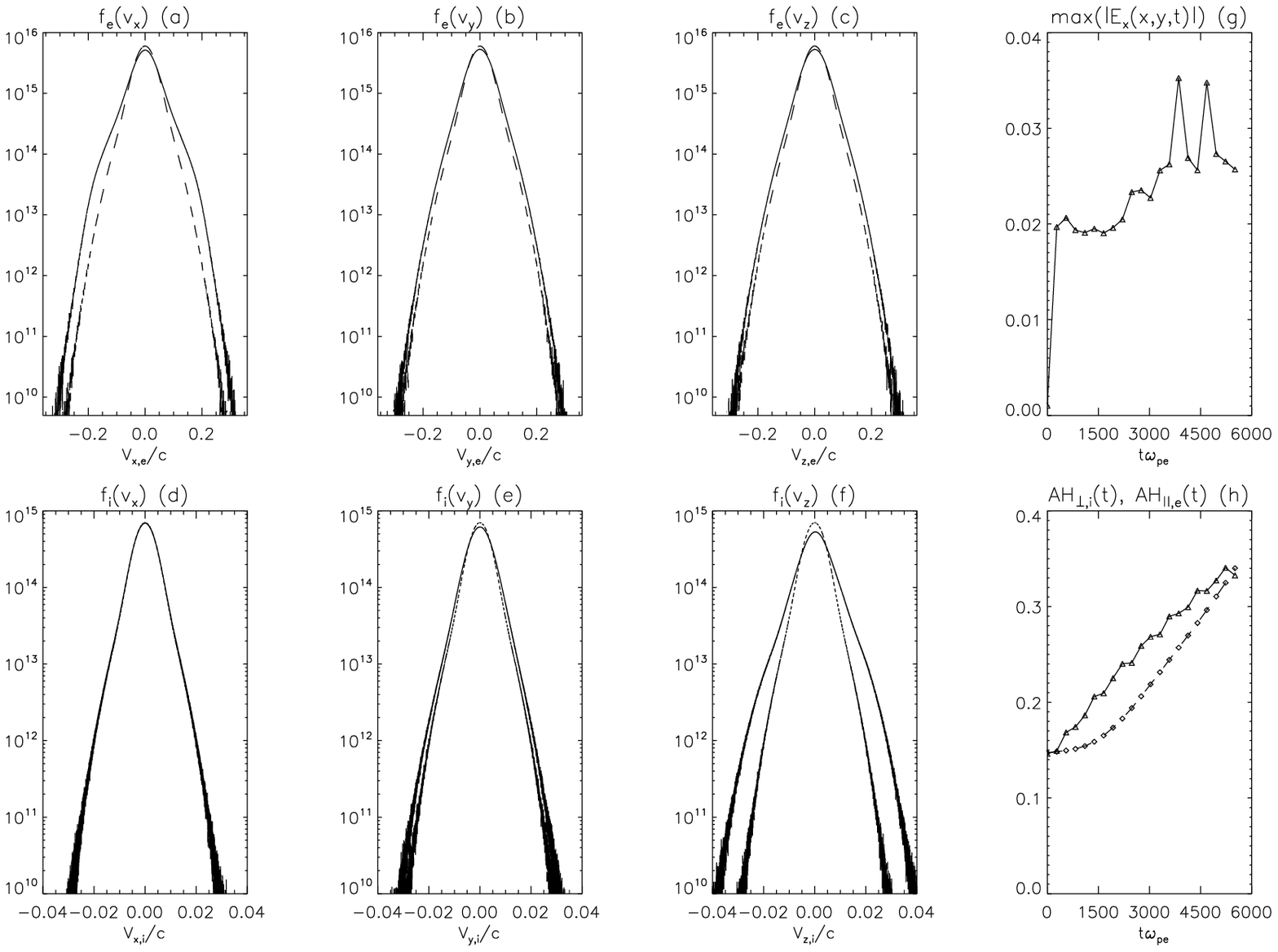}}
\caption{As in Fig.~3 but for the numerical run EL73.}
   \end{figure*}
   
We also gather from Figs.~11(a) and 11(b) that in the kinetic regime kink-like oscillation
amplitudes in the electron and ion density are larger  than in the inertial regime (cf. Figs.~1(a) and 1(b)).
Also, electron ion heating affects larger areas around the density gradients (Figs.~11(c) and 11(d)).

\begin{figure*}[htbp]   
\centerline{\includegraphics[width=0.65\textwidth]{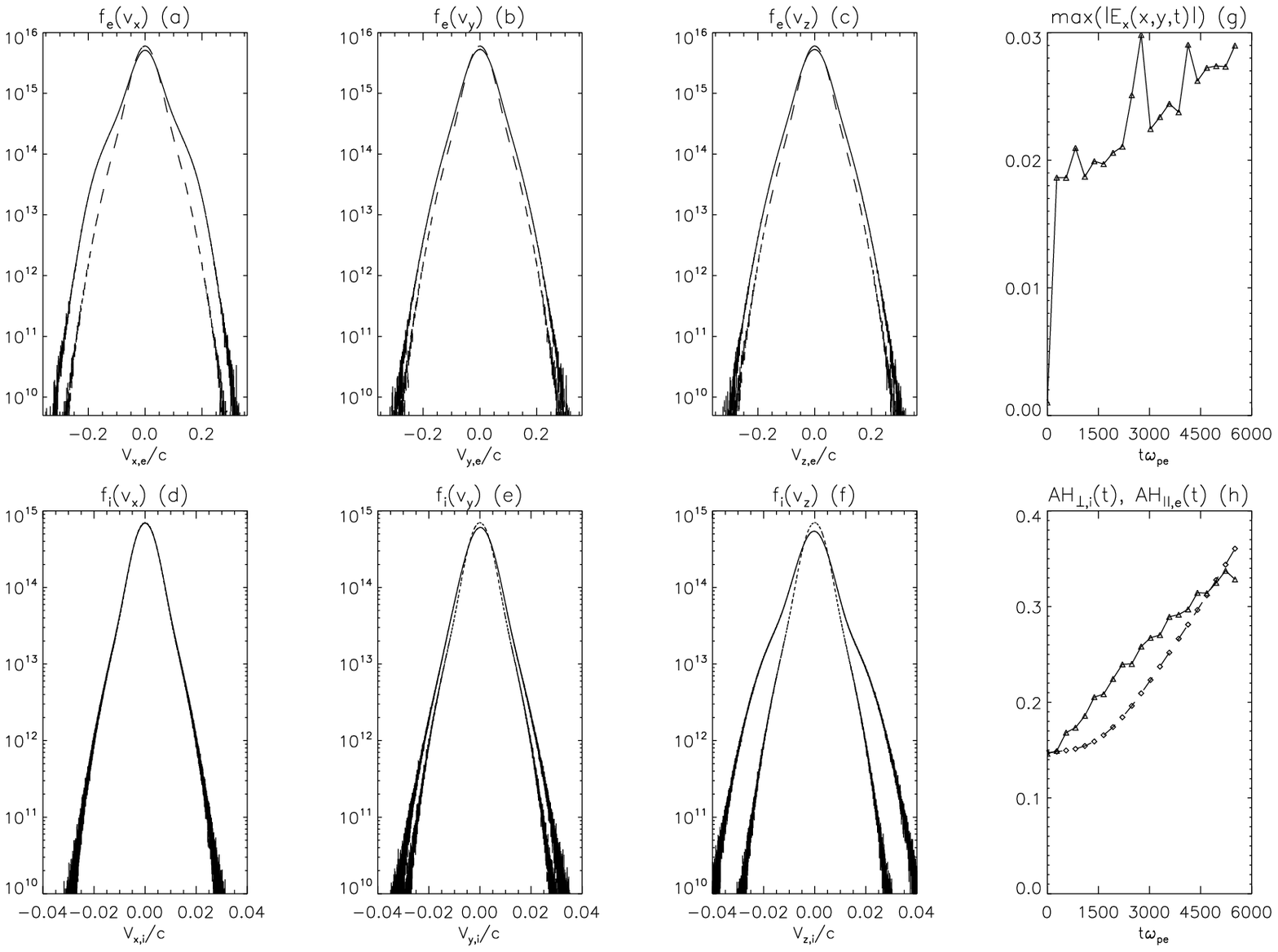}}
\caption{As in Fig.~3 but for the numerical run ER73.}
   \end{figure*}
Figs.~12, 13, 14 (kinetic regime) are analogous to    
Figs.~4, 5, 6 (inertial regime) and correspond to the runs R73, EL73 and ER73.
This means we can directly compare differences in the inertial 
and kinetic regimes. 
The conclusions that can be drawn are that 
in the kinetic  regime, R-circularly polarised, L-and R-elliptically polarised DAW with 
electric field in  the non-ignorable ($y$) transverse direction exceeding 6 times that of
in the ignorable ($z$) direction produce more prominent parallel  electron beams ($AH_{\parallel,e}(t)\approx 35\%$) and transverse ion beams
in the ignorable ($v_z$) direction, compared both to the corresponding runs in the inertial regime
and  L-circularly polarised wave in the kinetic regime.
Again, the electron beam speed coincides with the DAW speed.
Therefore the electron acceleration can again be understood
by a Landau damping of DAWs.
 
\begin{figure*}[htbp]
\centerline{\includegraphics[width=0.65\textwidth]{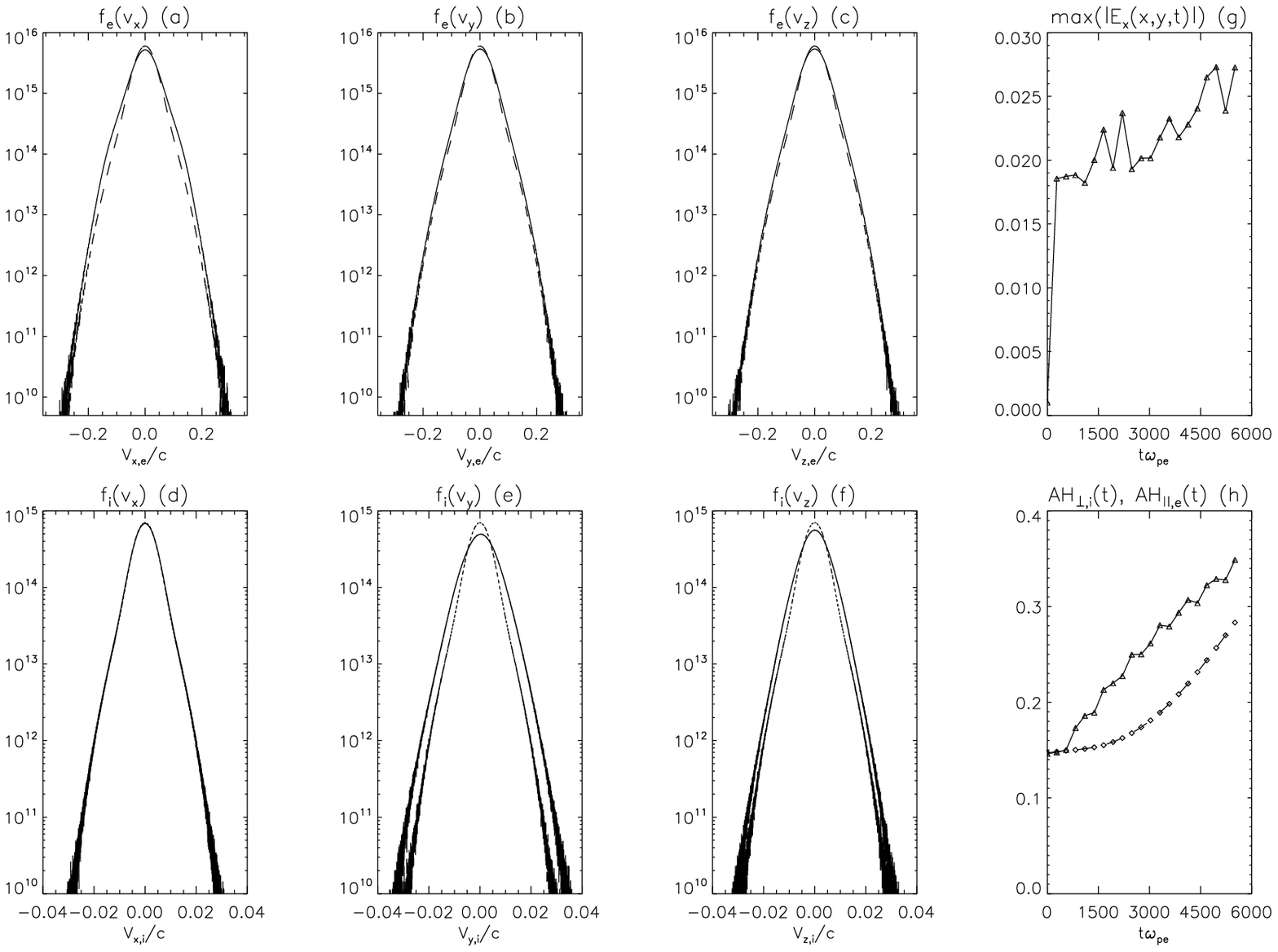}}
\caption{As in Fig.~3 but for the numerical run EL73$_1$.}
   \end{figure*}
Figs.~15 and 16 correspond to runs EL73$_1$ and ER73$_1$. 
We see from these figures that the behaviour is similar to run L73.  
L-and R-elliptically polarised DAW with 
electric field in  the non-ignorable ($y$) transverse direction being $1/6$th that of
in the ignorable ($z$) direction produce weak parallel electron beams and weak
transverse ion heating (no pronounced ion beams).
 
\begin{figure*}[htbp]
\centerline{\includegraphics[width=0.65\textwidth]{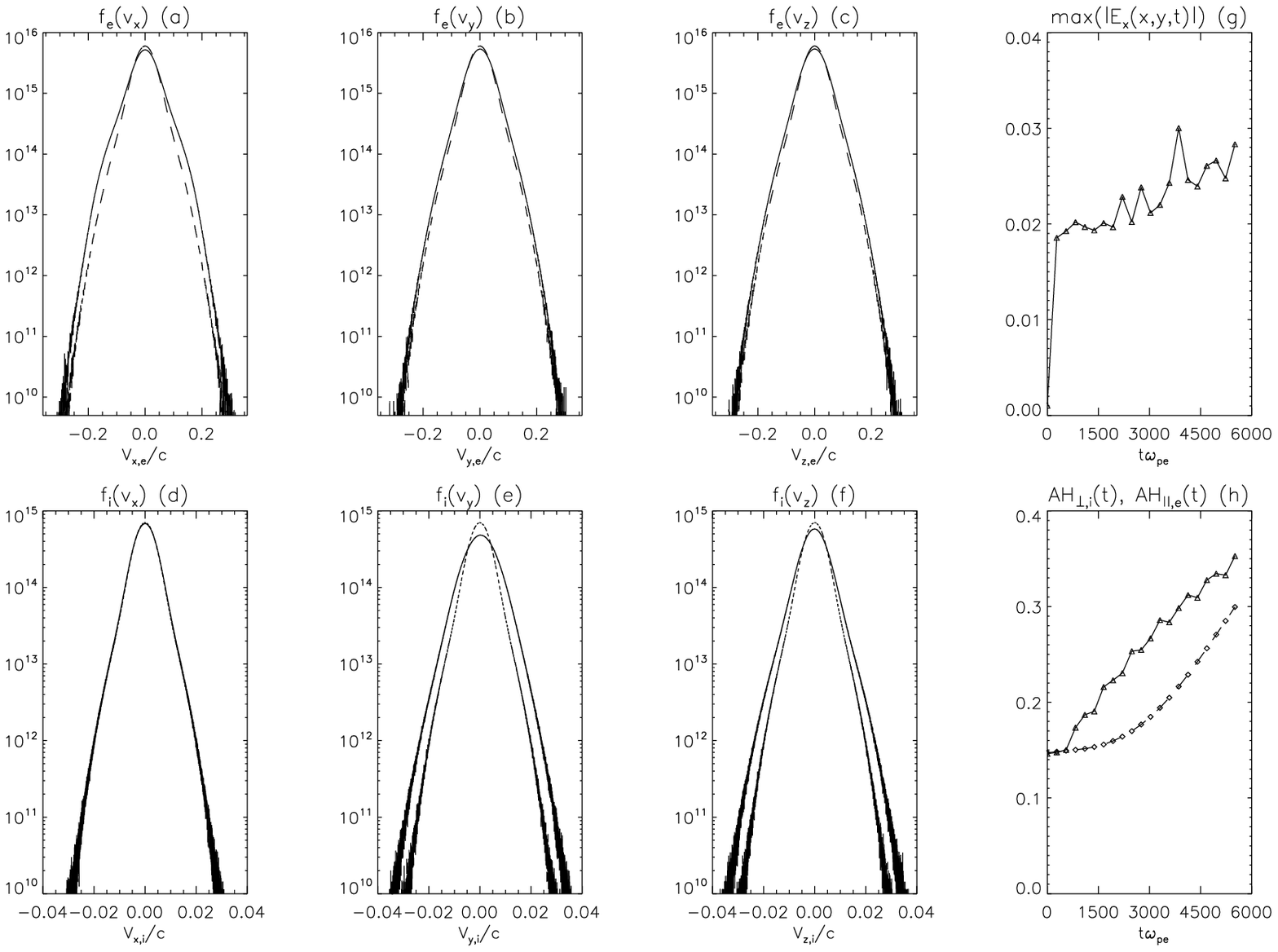}}
\caption{As in Fig.~3 but for the numerical run ER73$_1$.}
   \end{figure*}    

We finish presentation of the results by considering one long time
run L16Long which is similar to L16 but for 4 times larger $t_{end}=300\omega_{ci}^{-1}$.
Accordingly we had to increase the domain length in $x$-direction 4 times.
Thus in L16Long run we have $20000\times200$ grid size with $8\times10^8$ particles in the simulation. 
The purpose of run L16Long is to check whether numerical
simulation results presented in runs L16, R16, EL16, ER16, EL16$_1$ and ER16$_1$
have attained their respective "saturated" values, as for example
from Fig.~3(g) or Fig.~3(h) it is not entirely clear what happens 
to the amplitude of the generated parallel electric field and fraction
of accelerated particles in the density inhomogeneity regions for 
asymptotically large times. One has to realise that $t_{end}=75\omega_{ci}^{-1}$
is only $2.127\times 10^{-7}$ s (for mass ratio $m_i/m_e$).
Which is few orders of magnitudes smaller that impulsive stage
of the flare which is believed to be generating the considered DAWs.
   
\begin{figure*}
\centerline{\includegraphics[width=0.8\textwidth]{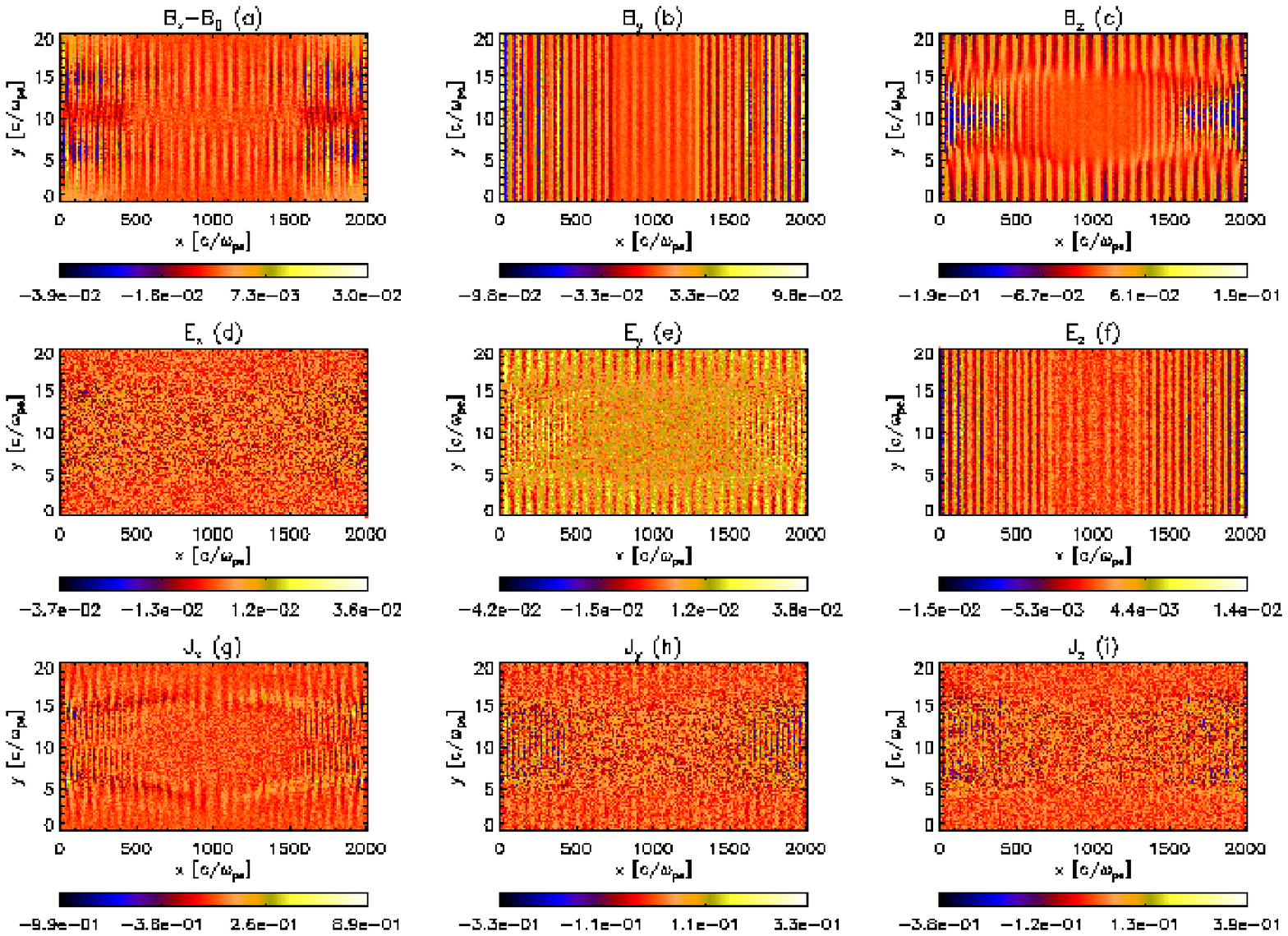}}
\caption{As in Fig.~1 but for the numerical run L16Long.}
\end{figure*}    

In Fig.~17 we plot time snapshots of various physical quantities at      
$t_{end}=300\omega_{ci}^{-1}$. Fig.~17 is analogous to Fig.~1 but for longer times
run. Commensurately we see larger number of full wavelength (namely 13)
developing in the driven DAW front. Again the simulation is stopped
before the DAWs collide, which Fig.~17 clearly demonstrates.
In Fig.~18 we show time evolution of the electron and ion distribution functions
(Figs.18(a)-(f)), amplitude of the generated $E_\parallel$ (Fig.~18(g)),
and $AH_{\parallel,e}(t)$ and $AH_{\perp,i}(t)$ indexes (Fig.~18(h)).
By comparison with Fig.~3, we gather from Fig.~18 that
parallel electron beam becomes more intense (Figs.~18(a)), and minor
transverse electron heating occurs too (Figs.~18(b),(c)).
Ion beams in the transverse direction form too but these are very weak.
It is interesting to note that 
electric field saturation level did not change much (still $\approx 0.03$)
both  $AH_{\parallel,e}(t)$ and $AH_{\perp,i}(t)$ indexes 
attain the same value of 30\%. Remeber that in case of L16 $AH_{\parallel,e}(t)$ attained
only 20\% by the time $t_{end}=75\omega_{ci}^{-1}$.
Thus we conclude that in run L16 the physical quantities have nearly reached
their asymptotic values.

\begin{figure*}[htbp]  
\centerline{\includegraphics[width=0.65\textwidth]{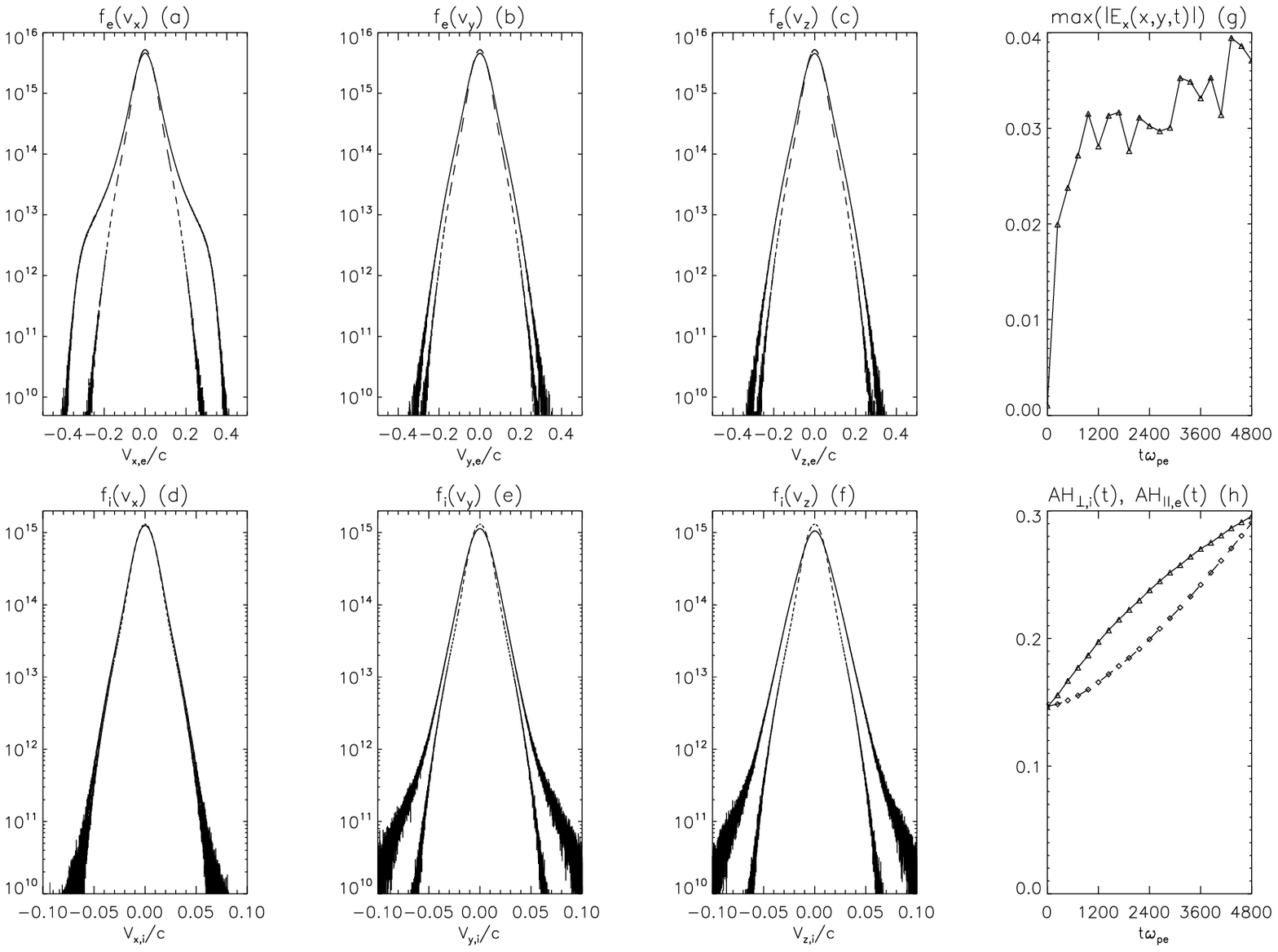}}
\caption{As in Fig.~3 but for the numerical run L16Long.}
\end{figure*}

\section{Conclusions}

X-ray observations of solar flares indicate that 
$10^{34}-10^{37}$ per second accelerated electrons are 
required to produce the observed hard X-ray emission. 
If the particle acceleration takes place
at the loop apex, this implies that for the
acceleration volume of $1-10$ Mm$^3$ with the number density of $n=10^{16}$ 
m$^{-3}$ all 100\% of electrons need to be accelerated.
Given the implausibility of the latter acceleration efficiency,
if the flare energy release triggers DAWs, these can rush
down the solar coronal loop towards the footpoints and accelerate electrons along the way to chromosphere
\cite{2008ApJ...675.1645F}. In this context 
and in the light of our initial study Ref.\cite{2005A&A...435.1105T},
here we study particle acceleration by the low frequency ($\omega=0.3\omega_{ci}$)
DAWs, which in fact are very similar to usual shear Alfven waves,
as far as their dispersion properties are concerned.
The DAWs propagate in the transversely inhomogeneous plasma.
In which density is increased in the middle, across the magnetic field, by factor of four and temperature
is lowered accordingly to keep the structure in pressure balance.
Such system mimics a solar coronal loop.
DAWs generate parallel electric field only in the density
inhomogeneity regions (loop edges). This parallel electric field behaviour coincides
with the parallel current  $\vec J = e (n_i \vec V_i - n_e \vec V_e)$.
The latter indicates that charge separation (due to different electron and ion
inertia) is the cause of the parallel electric field generation.
$E_\parallel$ then accelerates electrons (creates electron beams)
along the magnetic field,
while $E_\perp$ mostly heats ions in the transverse direction.    
In this study we focus on the effect of the wave polarisation, L- and R- circular and elliptical,
in the different regimes inertial ($\beta < m_e/m_i$) and kinetic 
($\beta > m_e/m_i$). 
Plasma beta is fixed at 0.02. Hence we vary the regimes by setting either
$m_e/m_i=1/16=0.0625 > \beta$ (inertial regime) and $m_e/m_i=1/73.44=0.0136 < \beta$ (kinetic regime).
The mass ratio of $73.44$ is not ideal (it corresponds to $1/25$th or the real mass ratio
of $1836$) but it is on the limit of available computational resources as a typical with $m_i/m_e=73.44$ run 
takes 48 hours on 256 processors (Here we present the results from total of 13 runs).
Our main goal was to study the latter effects on the 
efficiency of particle acceleration and the parallel electric field generation.
As a result we have established:
(i) In the inertial regime, fraction of accelerated electrons (along the magnetic field),
in the density gradient regions is $\approx 20\%$ by the time of IC driving of $t_{end}=75\omega_{ci}^{-1}$
which allows for the $3$ full wavelength to develop and  is increasing to 
$\approx 30\%$ by the time of $t_{end}=300\omega_{ci}^{-1}$ that allows for the $13$ wavelength to develop.
In all considered cases (see Table~\ref{runs}) ions are heated in the transverse to
the magnetic field direction and fraction of heated ions (with velocities above initial average thermal
speed) is $\approx 35 \%$.
(ii) While keeping the power of injected DAWs the same in all
considered numerical simulation runs, in the case of
right circular, left and right elliptical polarisation DAWs with the
electric field in the non-ignorable ($y$) transverse direction exceeding several times that of
in the ignorable ($z$) direction produce more pronounced parallel electron beams (with larger maximal
electron velocities) and transverse ion beams
in the ignorable ($v_z$) direction. 
In the inertial regime such polarisations yield the
fraction of accelerated electrons about  $20\%$.
In the kinetic regime this increases to $35\%$.
(iii) In the case of left circular, left and right elliptical polarisation DAWs with the
electric field in the non-ignorable ($y$) transverse direction several times less than that of
in the ignorable ($z$) direction produce less energetic parallel electron beams. 
In the inertial regime such polarisations yield the
fraction of accelerated electrons about  $20\%$.
In the kinetic regime this increases to $30\%$.
(iv) The parallel electric field that is generated in the density inhomogeneity
regions is independent of the regime (whether kinetic or inertial), i.e.
independent of electron-ion mass ratio and stays of the order $0.03\omega_{pe}c m_e /e$
which for considered solar flaring plasma parameters ($n=10^{16}$ m$^{-3}$ and $T=6\times10^7$K
at the domain edges) exceeds Dreicer electric field $E_D=n_e e^3 \ln \Lambda/(8 \pi \varepsilon_0^2k_BT)$
(with Coulomb logarithm $\ln \Lambda=17.75$) by a factor of $6.4 \times 10^8$.
Although our PIC simulations do not include particle collisions, and
thus Dreicer electric field is not directly relevant because this concept is
essentially collisional, the fact that the generated parallel electric field
exceeds Dreicer value by eight orders of magnitude indicates that
the accelerated electrons would be in the run-away regime.
The fact that in the present simulations 
$E_\parallel$ amplitude is independent of $m_i/m_e$ contradicts
with the scaling law ($E_\parallel \propto 1/(m_i/m_e)$) found by Ref.\cite{2007NJPh....9..262T}.
The difference can possibly explained by the fact that Ref.\cite{2007NJPh....9..262T}
considered cold plasma approximation i.e. all particles had the same velocity
(with the distribution function being a delta-function rather than a Maxwellian).
The latter would have prevented the possibility of having wave-particle interactions, which
as found in this study are rather important.
(v) Electron beam velocity has the phase velocity of the DAW.
This can be understood by  Landau damping of DAWs.
DAWs give up their energy to electrons via collisionless
Landau damping. Because for the
DAWs with $\omega=0.3 \omega_{ci}$ 
their phase seed is approximately the Alfven speed,
this, combined with the suggestion by Ref.\cite{2008ApJ...675.1645F}
that Alfven speed can be as high as few $\times 0.1c$,
for the realistic mass ratio of $m_i/m_e=1836$, would 
readily provide electrons with few tens of keV.
e.g. for $V_A=0.3c$  the electron energy would be 23 keV.
(vi) As we increased the mass ratio from $m_i/m_e=16$ to
73.44 the fraction of accelerated electrons has increased from $20\%$
to $30-35\%$ (depending on DAW polarisation).
This is because the velocity of the beam has shifted to
lower velocity (compare e.g. Fig.3(a) to 9(a)).
Since there are {\it always more electrons with a smaller velocity than higher 
velocity} in the Maxwellian distribution, 
for the mass ratio $m_i/m_e=1836$ the fraction of accelerated electrons
would be even higher than $35\%$. 
(vii) DAWs generate significant
density and temperature perturbations that are located in the density gradient regions
(solar coronal loop edges). These perturbations of density and temperature
should in principle be detectable in  {\it future}
observations because current time cadence of space instruments
is well above $\omega_{ci}^{-1}$.

The overall conclusion is that DAWs, propagating
in the transversely inhomogeneous plasma, 
in the Alfvenic ($\omega < \omega_{ci}$)
range of the plasma dispersion can effectively accelerate electrons
along the magnetic field and heat the ions in the transverse direction.
This is bound to have applications in many fields -- Earth
magnetosphere and laboratory plasma, in addition to the solar flare
context emphasised here.

\begin{acknowledgments}
The author would like to thank EPSRC-funded 
Collaborative Computational Plasma Physics  (CCPP) project
lead by Prof. T.D. Arber (Warwick) for providing 
EPOCH Particle-in-Cell code and Dr. K. Bennett (Warwick) for
CCPP related programing support. 
Computational facilities used are that of Astronomy Unit, 
Queen Mary University of London and STFC-funded UKMHD consortium at St Andrews
University. The author is financially supported by HEFCE-funded 
South East Physics Network (SEPNET).
\end{acknowledgments}


\end{document}